\documentclass[sigconf]{acmart}

\usepackage{setspace}
\usepackage{balance}

\AtBeginDocument{%
  \providecommand\BibTeX{{%
    \normalfont B\kern-0.5em{\scshape i\kern-0.25em b}\kern-0.8em\TeX}}}

\setcopyright{none}
\renewcommand\footnotetextcopyrightpermission[1]{}

\begin{document}

\title{Agile SoC Development with Open ESP}


\author{Paolo Mantovani, Davide Giri, Giuseppe Di Guglielmo, Luca Piccolboni,
  Joseph Zuckerman, \mbox{Emilio G. Cota}, Michele Petracca, Christian Pilato, and Luca P. Carloni}
\authornote{Emilio G. Cota is now with Google. Michele Petracca is now with
  Cadence Design Systems. Christian Pilato is now with Politecnico di Milano.}
\email{{paolo, davide_giri, giuseppe, piccolboni, jzuck, cota, petracca, pilato, luca}@cs.columbia.edu}
\affiliation{%
  \institution{Department of Computer Science, Columbia University in the City of New York, New York, NY 10027}
}

\renewcommand{\shortauthors}{Mantovani, et al.}

\begin{abstract}
  ESP is an open-source research platform for heterogeneous SoC
  design.
  The platform combines a modular tile-based architecture with a variety of
  application-oriented flows for the design and optimization of
  accelerators. The ESP architecture is highly scalable and strikes a balance
  between regularity and specialization. The companion methodology raises the
  level of abstraction to system-level design and enables an automated flow from
  software and hardware development to full-system prototyping on FPGA.
  For application developers, ESP offers domain-specific automated solutions to
  synthesize new accelerators for their software and to map complex workloads
  onto the SoC architecture. For hardware engineers, ESP offers
  automated solutions to integrate their accelerator designs into the complete
  SoC. Conceived as a heterogeneous integration platform and tested through
  years of teaching at Columbia University, ESP supports the open-source
  hardware community by providing a flexible platform for agile SoC development.
\vspace{-0.2cm}
\end{abstract}

\keywords{System-level design, SoC, accelerators, network-on-chip.}


\maketitle
\pagestyle{plain}

\section{Introduction}
\label{sec:intro}

\textbf{Why ESP?} ESP is an open-source research platform for heterogeneous
system-on-chip (SoC) design and programming~\cite{esp}.
ESP is the result of nine years of research and teaching at Columbia
University~\cite{carloni_dac16,carloni_wcae19}.
Our research was and is motivated by the consideration that Information Technology
has entered the age of heterogeneous computing.
From embedded devices at the edge of the cloud to data center blades at the core
of the cloud, specialized hardware accelerators are increasingly employed to
achieve energy-efficient performance~\cite{borkar11,Horowitz2014,caulfield2016}.
Across a variety of application domains, such as mobile electronics, automotive,
natural-language processing, graph analytics and more, computing systems rely on
highly heterogeneous SoC architectures.
These architectures combine general-purpose processors with a variety of
accelerators specialized for tasks like image processing, speech
recognition, radio communication and graphics~\cite{dally2020} as well as
special-purpose processor cores with custom instruction sets, graphics
processing units, and tensor manipulation units~\cite{jouppi2018}.
The shift of the silicon industry from homogeneous multicore processors to
heterogeneous SoCs is particularly noticeable if one looks at the portion of
chip area dedicated to accelerators in subsequent generations of
state-of-the-art chips for smartphones~\cite{shao_ieeemicro_15},
or at the amount of diverse processing elements in chips for autonomous
driving~\cite{chishiro_ICESS19}.

\begin{figure}[t]
  \centering
  \includegraphics[width=0.9\columnwidth]{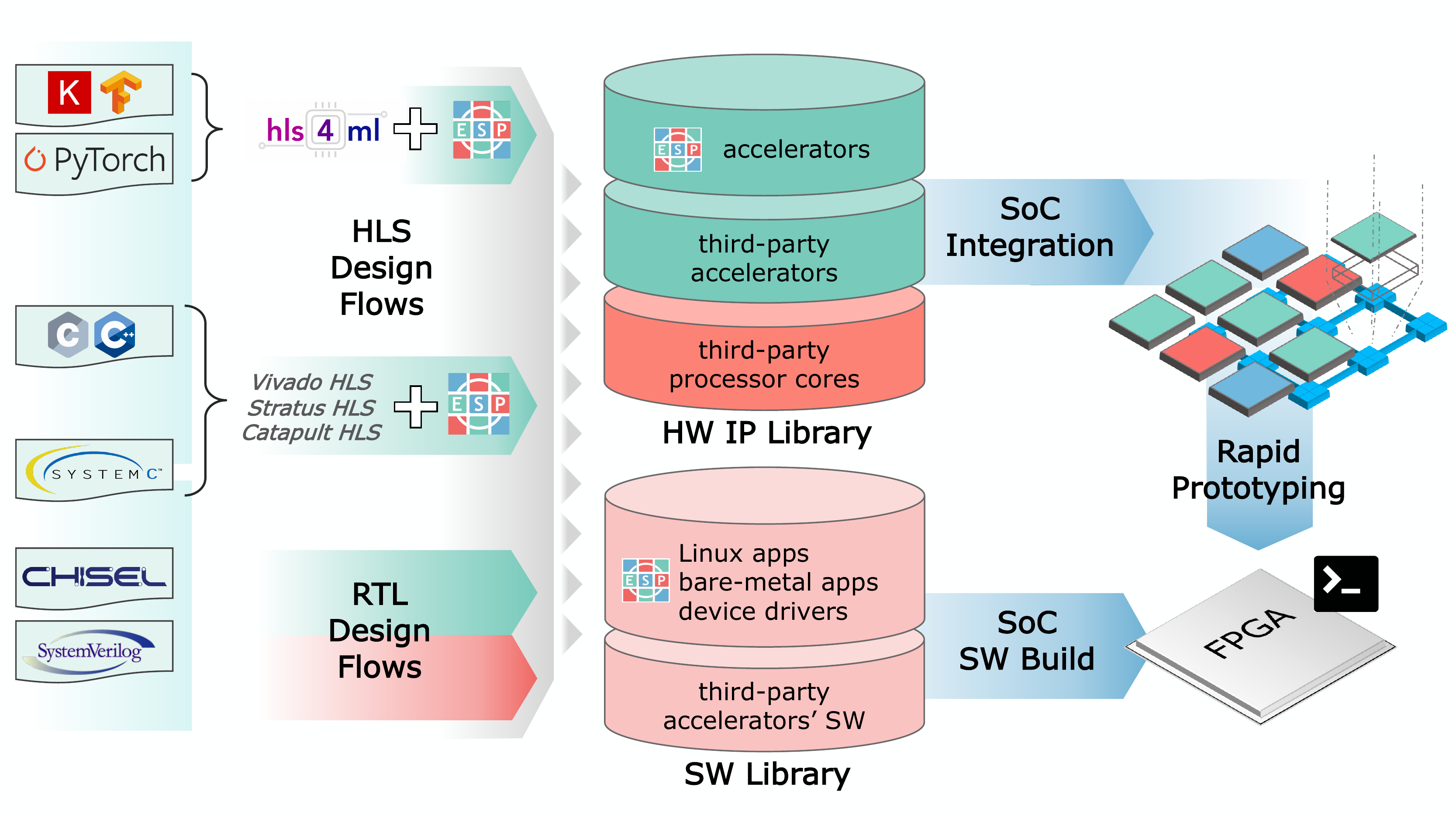}
  \vspace{-0.1in}
  \caption{Agile SoC design and integration flows in ESP.}
  \label{fig:esp-overview}
  \vspace{-0.1in}
\end{figure}

\textbf{ESP Vision.} ESP is a platform, i.e., the combination of an
architecture and a methodology~\cite{carloni_dac16}.
The methodology embodies a set of agile SoC design and integration flows, as
shown in \figurename~\ref{fig:esp-overview}.
The ESP vision is to allow application domain experts to design SoCs.
Currently, ESP allows SoC architects to rapidly implement FPGA-based prototypes of complex SoCs.
The ESP \emph{scalable architecture} and its \emph{flexible methodology} enable
a seamless integration of third-party open-source hardware (OSH) components
(e.g., the Ariane RISC-V core~\cite{ariane,ariane_paper} or the
NVIDIA Deep-Learning Accelerator~\cite{nvdla}).
SoC architects can instantiate also accelerators that are developed with one of
the many design flows and languages supported by ESP.
The list, which continues to grow, currently includes: C/C++ with Xilinx Vivado
HLS and Mentor Catapult HLS; SystemC with Cadence Stratus HLS; Keras TensorFlow,
PyTorch and ONNX with hls4ml; and Chisel, SystemVerilog, and VHDL
for register-transfer level (RTL) design.
Hence, accelerator designers can choose the abstraction level and specification language
that are most suitable for their coding skills and the target computation kernels.
These design flows enable the creation of a rich library of components ready to
be instanced into the ESP tile-based architecture with the help of the SoC
integration flow.

Thanks to the automatic generation of device drivers from pre-designed templates,
the ESP methodology simplifies the invocation of accelerators from user-level
applications executing on top of Linux~\cite{giri_date20, mantovani_cases16}.
Through the automatic generation of a network-on-chip (NoC) from a parameterized
model, the ESP architecture can scale to accommodate many processors, tens of
accelerators, and a distributed memory hierarchy~\cite{giri_nocs18}.
A set of {\em platform services} provides pre-validated solutions to access or
manage SoC resources, including accelerators configuration~\cite{mantovani_aspdac16},
memory management~\cite{mantovani_cases16}, and dynamic voltage
frequency scaling (DVFS)~\cite{mantovani_dac16}, among others.
ESP comes with a GUI that guides the designers through the interactive
choice and placement of the tiles in the SoC and it has push-button
capabilities for rapid prototyping of the SoC on FPGA.

\textbf{Open-Source Hardware.} OSH holds the
promise of boosting hardware development and creating new opportunities for
academia and entrepreneurship~\cite{gupta17}. In recent years, no other project
has contributed to the growth of the OSH movement more than
RISC-V~\cite{asanovic_case14}.
To date, the majority of OSH efforts have focused on the development of
processor cores that implement the RISC-V ISA and small-scale SoCs that connect
these cores with tightly-coupled functional units and coprocessors, typically
through bus-based interconnects.  Meanwhile, there have been less efforts in
developing solutions for large-scale SoCs that combine RISC-V cores with many
loosely-coupled components, such as coarse-grained accelerators~\cite{cota_dac15},
interconnected with a NoC. With this gap in mind, we have made an open-source
release of ESP to provide the OSH community with a platform for heterogeneous
SoC design and prototyping~\cite{esp}.

\section{The ESP Architecture}
\label{sec:arch}

\begin{figure}[t]
  \begin{center}
    \resizebox{0.95\columnwidth}{!}{\includegraphics{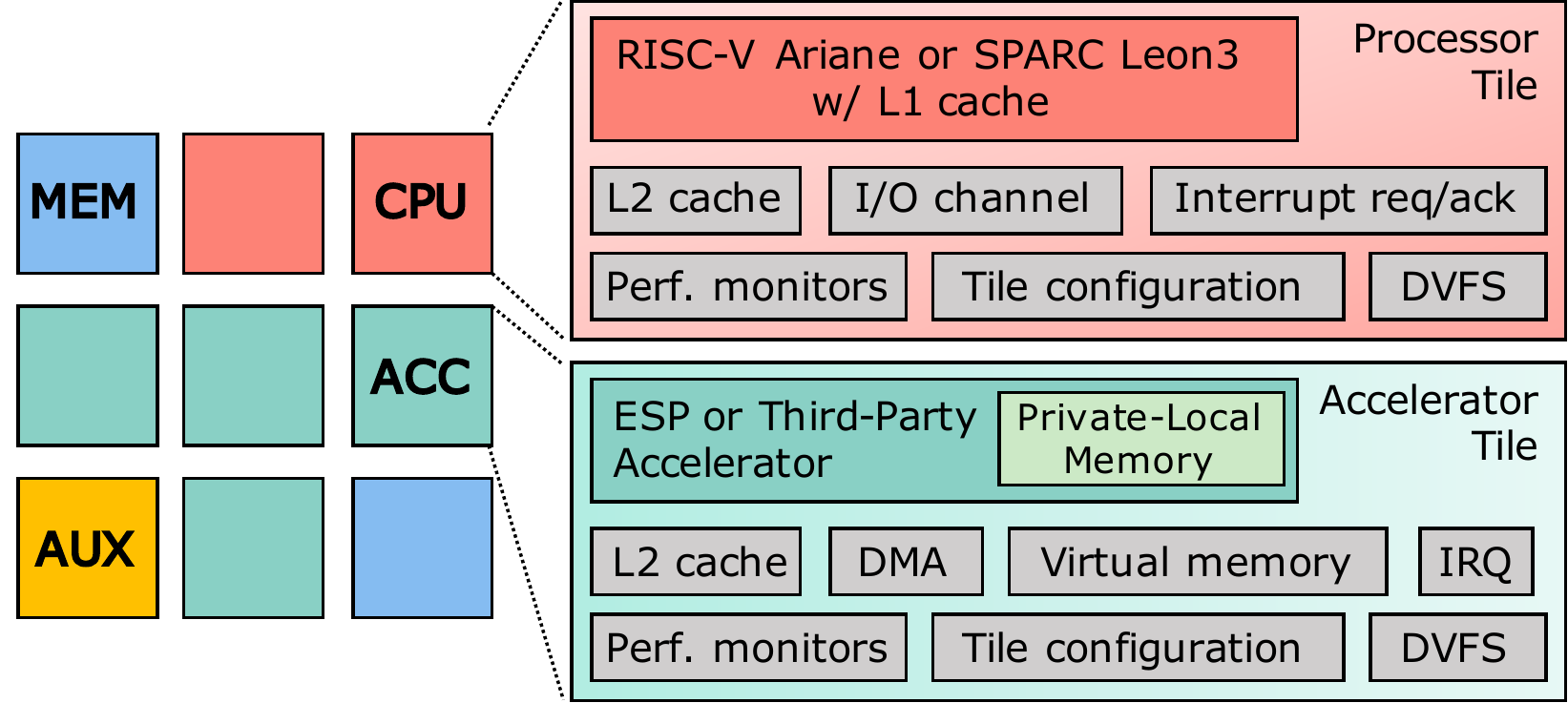}}
  \end{center}
  \vspace{-0.1in}
  \caption{Example of a 3x3 instance of ESP with a high-level overview of the
    sockets for processors and accelerators. }
\label{fig:esp-arch}
\end{figure}

The ESP architecture is structured as a heterogeneous tile grid. For a given
application domain, the architect decides the structure of the SoC by
determining the number and mix of tiles.
For example, \figurename~\ref{fig:esp-arch} shows a 9-tile SoC organized in a
$3 \times 3$ matrix.
There are four types of tiles: processor tile, accelerator tile, memory tile for
the communication with main memory, and auxiliary tile for peripherals, like UART
and Ethernet, or system utilities, like the interrupt controller and the timer.
To support a high degree of scalability, the ESP tiles are connected by a
{\em multiplane NoC}~\cite{yoon_tcad13}.

The content of each tile is encapsulated into a {\em modular socket}
(aka {\em shell}), which interfaces the tile to the NoC and implements the
platform services.
The socket-based approach, which decouples the design of a tile from the
design of the rest of the system, is one of the key elements of the agile ESP
SoC design flow. It highly simplifies the design effort of each tile by taking
care of all the system integration aspects, and it facilitates the reuse of
intellectual property (IP) blocks.
For instance,
the ESP accelerator socket implements services for
DMA, cache coherence, performance monitors,
and distributed interrupt requests.

At design time, it is possible to choose the set of services to instantiate in
each tile. At runtime, the services can be enabled and many of them offer
reconfigurability options, e.g., dynamic reconfiguration of the cache-coherence
model~\cite{giri_aspdac19}.

The ESP architecture implements a distributed system that is inherently
scalable, modular and heterogeneous, where processors and accelerators are given
the same importance in the SoC.
Differently from other OSH platforms, ESP proposes a system-centric view, as
opposed to a processor-centric view.

\begin{figure*}[t]
  \begin{center}
    \resizebox{1\textwidth}{!}{\includegraphics{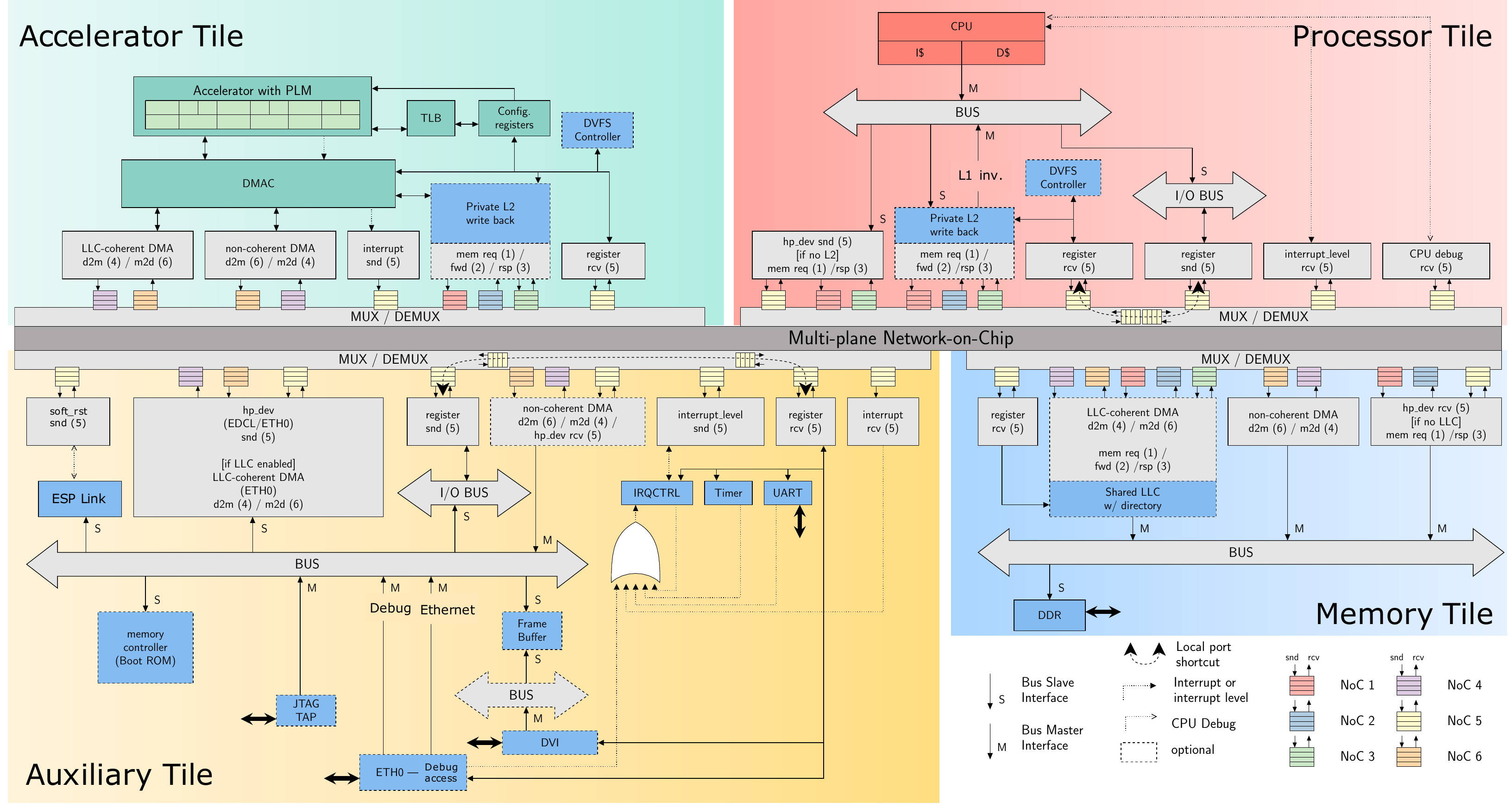}}
  \end{center}
  \vspace{-0.1in}
  \caption{Detailed architecture of the NoC interface for four main ESP tiles.
  }
  \label{fig:details}
\end{figure*}

\vspace{-0.1in}
\subsection{System Interconnect}
Processing elements act as transaction masters that access peripherals
and slave devices distributed across remote tiles. All remote communication is
supported by a NoC, which is a transparent communication layer.
Processors and accelerators, in fact, operate as if all remote components were
connected to their local bus controller in the ESP sockets. The sockets include
standard bus ports, bridges, interface adapters and proxy components that
provide a complete decoupling from the network interface.
\figurename~\ref{fig:details} shows in detail the modular architecture of the
ESP interconnect for the case of a six-plane NoC. Every platform service is
implemented by a pair of proxy components. One proxy translates requests from
bus masters, such as processors and accelerators, into transactions for one of
the NoC planes. The other proxy forwards requests from the NoC planes to the
target slave device, such as last-level cache (LLC) partitions, or Ethernet.
For each proxy, there is a corresponding buffer queue, located between the tile
port of the NoC routers and the proxy itself.
In \figurename~\ref{fig:details}, the color of a queue depends on the assigned
NoC plane. The number and direction of the arrows connected to the queues
indicate whether packets can flow from the NoC to the tile, from the tile to the
NoC, or in both directions. The arrows connect the queues to the proxies.
These are labeled with the name of the services they implement and the number of
the NoC plane used for receiving and sending packets, respectively.

The current implementation of the ESP NoC is a packet-switched 2D-mesh topology
with look-ahead dimensional routing.
Every hop takes a single clock cycle because arbitration and next-route
computation are performed concurrently.
Multiple physical planes allow protocol-deadlock prevention and provide
sufficient bandwidth for the various message types.
For example, since a distributed directory-based protocol for cache coherence
requires three separate channels,
planes 1, 2 and 3 in \figurename~\ref{fig:details} are assigned to request,
forward, and response messages, respectively.
Concurrent DMA transactions, issued by multiple accelerators and handled by
various remote memory tiles, require separate request and response
planes. Instead of reusing the cache-coherence planes, the addition of two new
planes (4 and 6 in \figurename~\ref{fig:details}) increases the overall NoC bandwidth.
Finally, one last plane is reserved for short messages, including interrupt,
I/O
configuration, monitoring and debug.

Currently, customizing the NoC topology is not automated in the ESP SoC integration flow.
System architects, however, may explore different topologies
by modifying the router instances
and updating the logic to generate the header flit for the NoC packets~\cite{yoon_nocs17}.

\vspace{-0.1in}
\subsection{Processor Tile} \label{sec:cpu-tile}
Each processor tile contains a processor core that is chosen at design time
among those available: the current choice is between the RISC-V 64-bit Ariane
core from ETH Zurich~\cite{ariane,ariane_paper} and the SPARC 32-bit LEON3 core
from Cobham Gaisler~\cite{leon3}.
Both cores are capable of running Linux and they come with their private L1
caches. The processor integration into the distributed ESP system is
transparent: no ESP-specific software patches are needed to boot Linux.
Each processor communicates on a local bus and is agnostic of the rest of the
system. The memory interface of the LEON3 core requires a 32-bit AHB bus,
whereas Ariane comes with a 64-bit AXI interface.
In addition to proxies and bus adapters, the processor socket provides a unified
private L2 cache of configurable size, which implements a directory-based MESI
cache-coherence protocol.
Processor requests directed to memory-mapped I/O registers are forwarded by the
socket to the IO/IRQ NoC plane through an APB adapter.
The only processor-specific component of the socket is an interrupt-level proxy,
which implements the custom communication protocol between the processor and the
interrupt controller and system timer in the auxiliary tile.

\vspace{-0.1in}
\subsection{Memory Tile} \label{sec:mem-tile}
Each memory tile contains a channel to external DRAM.  The number of memory
tiles can be configured at design time. Typically, it varies from one to four
depending on the size of the SoC.
All necessary hardware logic to support the partitioning of the addressable
memory space is automatically generated and the partitioning is completely
transparent to software.
Each memory tile also contains a configurably-sized partition of the LLC with
the corresponding directory.
The LLC in ESP implements an extended MESI protocol, in combination with the
private L2 cache in the processor tiles, that supports Linux with symmetric
multiprocessing, as well as runtime reconfigurable coherence for
accelerators~\cite{giri_aspdac19}.

\subsection{Accelerator Tile} \label{sec:acc-tile}
This tile contains the specialized hardware of a loosely-coupled accelerator~\cite{cota_dac15}.
This type of accelerator executes a coarse-grained task independently from the
processors while exchanging large datasets with the memory hierarchy.
To be integrated in the ESP tile, illustrated on the top-left portion of
\figurename~\ref{fig:details}, accelerators should comply to a simple interface
that includes load/store ports for latency-insensitive
channels~\cite{carloni_iccad99,carloni_pieee15},
signals to configure and start the accelerator, and an {\small\texttt{acc\_done}}
signal to notify the accelerator completion and generate an interrupt for the processors.
ESP accelerators that are newly designed with one of the supported design flows
automatically comply with this interface.
For existing accelerators, ESP offers a third-party integration flow~\cite{giri_carrv20}.
In this case, the accelerator tile has only a subset of the proxy components
because the configuration registers, DMA for memory access, and TLB for virtual
memory~\cite{mantovani_cases16} are replaced by standard bus adapters.

The set of platform services provided by the socket relieves the designer from
the burden of ``reinventing the wheel'' with respect to implementing accelerator
configuration through memory-mapped registers, address translation, and
coherence protocols. Furthermore, the socket enables point-to-point
communication (P2P) among accelerator tiles so that they can exchange data
directly instead of using necessarily shared-memory communication.

Third-party accelerators can use the services to issue interrupt requests,
receive configuration parameters and initiate DMA transactions.
They are responsible, however, for
managing shared resources, such as reserved memory regions, and for implementing
their own custom hardware-software synchronization protocol.

The run-time reconfigurable coherence protocol service is particularly relevant for
accelerators. In fact, there is no
static coherence protocol that can necessarily serve well all invocations of a
set of heterogeneous accelerators in a given SoC~\cite{giri_ieeemicro18}.
With the \emph{non-coherent DMA} model, an accelerator  bypasses the cache
hierarchy to exchange data directly with main memory.
With the \emph{fully-coherent} model, the accelerator communicates with an
optional private cache placed in the accelerator socket. The ESP cache hierarchy
augments a directory-based MESI protocol with support for two models where
accelerators send requests directly to the LLC, without owning a private cache:
the \emph{LLC-coherent DMA} and the \emph{coherent DMA} models. The latter keeps
the accelerator requests coherent with respect to all private caches in the
system, whereas the former does not. While \emph{fully-coherent} and
\emph{coherent DMA} are fully handled in hardware by the ESP cache
hierarchy, \emph{non-coherent DMA} and \emph{LLC-coherent DMA} demand that
software acquires appropriate locks and flushes private caches before invoking accelerators.
These synchronization mechanisms are implemented by the ESP device
drivers, which are generated automatically
when selecting any of the supported HLS flows discussed in Section~\ref{sec:method}.

\vspace{-0.1in}
\subsection{Auxiliary Tile} \label{sec:aux-tile}
The auxiliary tile hosts all shared peripherals in the system except from
memory: the Ethernet NIC, UART, a digital video interface, a debug link to
control ESP prototypes on FPGA and a monitor module that collects various
performance counters and periodically forwards them through the Ethernet
interface.

As shown in \figurename~\ref{fig:details}, the socket of the auxiliary tile is
the most complex because most platform services must be available to serve the
devices hosted by this tile. The interrupt-level proxy, for instance, manages
the communication between the processors and the interrupt controller.
Ethernet, which requires coherent DMA to operate as a slave peripheral, enables
users to remotely log into an ESP instance via SSH. The frame-buffer memory,
dedicated to the video output, is connected to one proxy for memory-mapped I/O
and one for non-coherent DMA transactions. These enable both processor cores and
accelerators to write directly into the video frame buffer.
The Ethernet debug interface~\cite{grlib}, instead, uses the memory-mapped I/O
and register access services to allow ESP users to monitor and debug the system
through the \emph{ESP Link} application. Symmetrically, UART, timer, interrupt
controller and the bootrom are controlled by any master in the system through
the counterpart proxies for memory-mapped I/O and register access. Hence, the
auxiliary tile includes both pairs of proxies. These enable an additional
communication path, labeled as \emph{local port shortcut} in
\figurename~\ref{fig:details}, which connects the masters in the auxiliary tile
(i.e. the Ethernet debug link) with slaves that do not share the same local bus.
A similar shortcut in the processor tile allows a processor to flush its own
private L2 cache and manage its local DVFS controller.

\section{The ESP Software Stack}
\label{sec:software}

\begin{figure}[t]
  \centering
  \includegraphics[width=1.0\columnwidth]{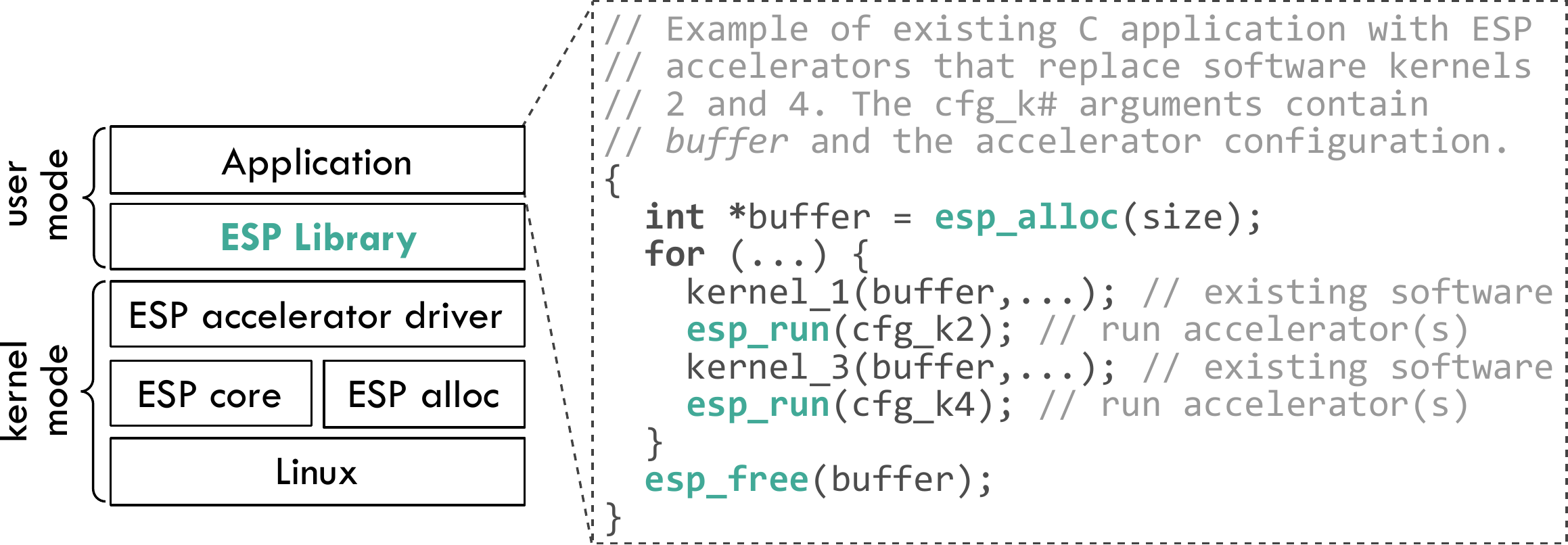}
  \vspace{-0.2in}
  \caption{ESP accelerator API for seamless shared memory.}
  \label{fig:software}
  \vspace{-0.1in}
\end{figure}

The ESP accelerator's Application Programming Interface (API) library simplifies
the invocation of accelerators from a user application, by exposing only three
functions to the programmer~\cite{giri_date20}.
Underneath, the API invokes the accelerators with the automatically generated
Linux device drivers.
The API is lightweight and can be targeted from existing applications or by a
compiler.
For a given application, the software execution of a computationally intensive
kernel can be replaced with hardware accelerators by means of a single function
call ({\small \texttt{esp\_run()}}).
\figurename~\ref{fig:software} shows the case of an application with four
computation kernels, two executed in software and two implemented with an
accelerator. The configuration argument passed to {\small \texttt{esp\_run()}}
is a simple data structure that specifies which accelerator(s) to invoke, how to
configure them, and their point-to-point dependencies, if any.
By using the {\small \texttt{esp\_alloc()}} and {\small \texttt{esp\_free()}}
functions for memory allocation, data can be truly shared between accelerators and processors,
i.e., no data copies are necessary. Data are allocated in an efficient way
to improve the accelerator's access to memory without compromising the
software's performance~\cite{mantovani_cases16}.
The ESP software stack, combined with the generation of device drivers for new
custom accelerators, makes the accelerator invocation as transparent as possible
for the application programmers.

\section{The ESP Methodology}
\label{sec:method}

The ESP design methodology is flexible because it embodies different design
flows, for the most part automated and supported by commercial CAD tools.
In particular, recalling \figurename~\ref{fig:esp-overview},
the {\em accelerator design flow} (on the left in the figure) aids the creation
of an IP library, whereas the {\em SoC flow} (on the right) automates the
integration of heterogeneous components into a complete SoC.

\subsection{Accelerator Flow}
The end goal of this flow is to add new elements to the library of accelerators
that can be automatically instantiated with the SoC flow.
Designers can work at different abstraction levels with various specification
languages:

\begin{itemize}
\item Cycle-accurate RTL descriptions with languages like VHDL, Verilog,
  SystemVerilog, or Chisel.
\item Loosely-timed or un-timed behavioral descriptions with SystemC or C/C++
  that get synthesized into RTL with high-level synthesis (HLS) tools. ESP
  currently supports the three main commercial HLS tools: Cadence Stratus HLS,
  Mentor Catapult, and Xilinx Vivado HLS.
\item Domain-specific libraries for deep learning like Keras TensorFlow,
  PyTorch, and ONNX, for which ESP offers a flow combining HLS tools with
  hls4ml, an OSH project~\cite{hls4ml,duarte_2018}.
\end{itemize}

\textbf{HLS-Based Accelerator Design.}
For the HLS-based flows, ESP facilitates the job of the accelerator designer by
providing ESP-compatible accelerator templates, HLS-ready skeleton
specifications, multiple examples, and step-by-step tutorials for each flow.

The push in the adoption of HLS from C/C++ specifications has many reasons:
(1) a large codebase of algorithms written in these languages,
(2) a simplified hardware/software co-design (since most embedded software is
in C), and
(3) a thousand-fold faster functional execution of the specification than the
counterpart RTL simulation.
On the other hand, HLS from C/C++ has also shown some limitations because of the
lack of ability to specify or accurately infer concurrency, timing, and
communication properties of the hardware systems.
HLS flows based on the IEEE-standard language SystemC overcome these
limitations, thus making SystemC the de-facto standard to model protocols and
control-oriented applications at a level higher than RTL.

In ESP, we support and encourage the use of both C/C++ and SystemC flows for
HLS and we have defined a set of guidelines to support the designers in porting
an application to an HLS-ready format.
The ideal starting point is a self-contained description of the computation
kernel, written in a subset of the C/C++ language~\cite{nane16}:
a limited use of pointers and the absence of dynamic memory allocation and
recursion are important; also, aside from common mathematical functions,
no external library functions should be used.
This initial software transformation is oftentimes the most important step to
obtain good quality source code for HLS~\cite{mantovani_aspdac16}.

\begin{figure}[t]
  \centering
  \includegraphics[width=1.0\columnwidth]{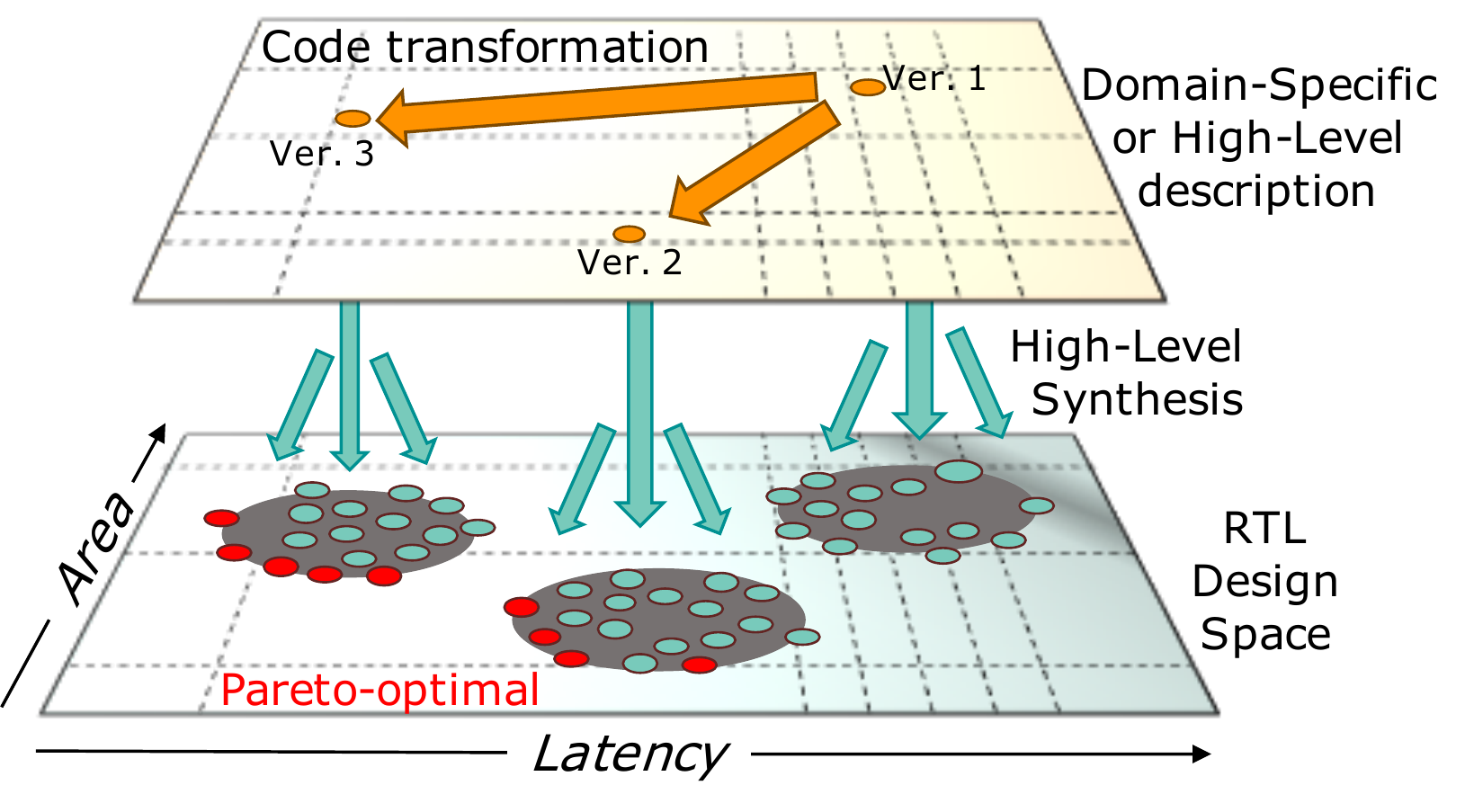}
    \vspace{-0.2in}
    \caption{HLS-based accelerator design in ESP.}
  \label{fig:dse}
\end{figure}

The designer of an ESP accelerator should aim at a well-structured description that
partitions the specification into concurrent functional blocks.
The goal is to obtain any synthesizable specification that enables the
exploration of a vast design space, by evaluating many micro-architectural and
optimization choices.
\figurename~\ref{fig:dse} shows the relationship between the C/C++/SystemC
design space and the RTL design space. The HLS tools provide a rich set of
configuration knobs to obtain a variety of RTL implementations, each
corresponding to a different cost-performance tradeoff point~\cite{liu_date12,liu_dac13}.
The knobs are push-button directives of the HLS tool represented by the green arrows.
Designers may also perform manual transformations of the specification (orange
arrows) to explore the design space while preserving the functional behavior.
For example, they can expose parallelism by removing false dependencies or they
can reduce resource utilization by encapsulating sections of code with similar
behavior into callable functions~\cite{mantovani_aspdac16}.

\begin{figure}[t!]
  \centering
      \includegraphics[width=0.6\columnwidth]{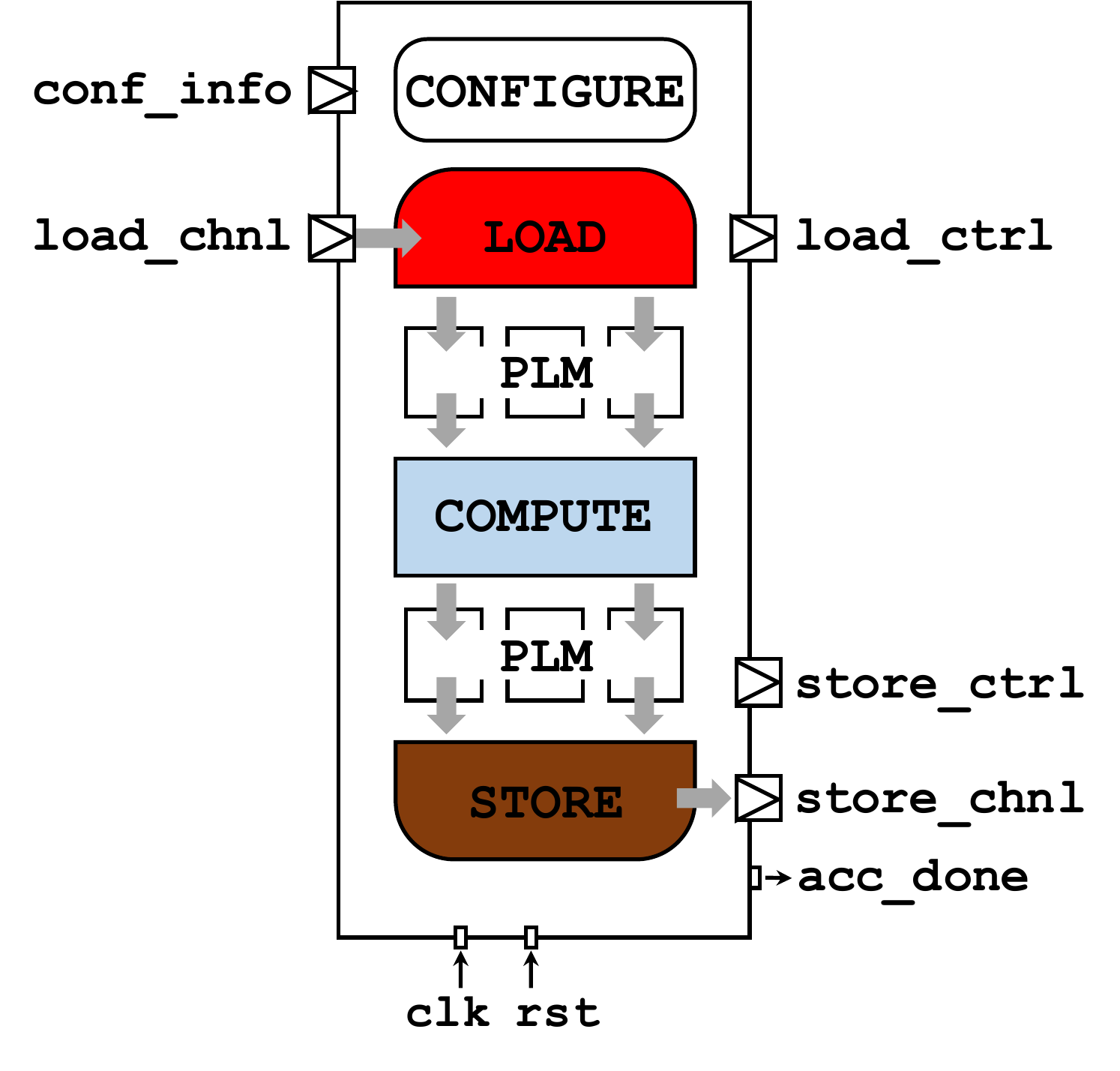}
    \vspace{-0.1in}
      \caption{Structure of ESP accelerators.}
     \vspace{-0.2in}
      \label{fig:accelerator-struct}
\end{figure}

\begin{figure*}[t!]
  \centering
      \includegraphics[width=0.9\textwidth]{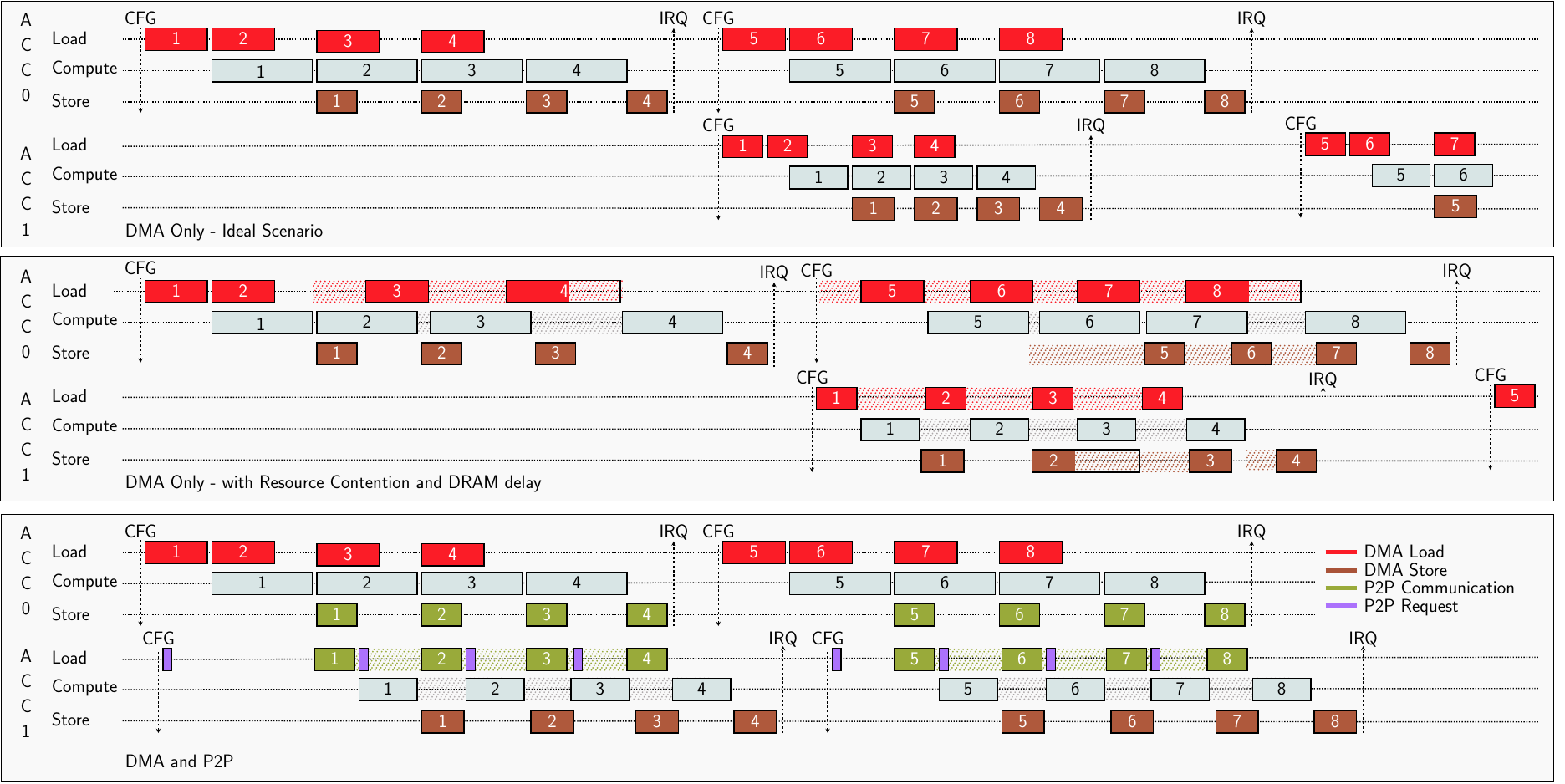}
      \caption{Overlapping of computation and communication of ESP accelerators.}
      \label{fig:accelerator-comm}
\end{figure*}

\textbf{Accelerator Structure.}
The ESP accelerators are based on the loosely-coupled model~\cite{cota_dac15}.
They are programmed like devices by applications that invoke device drivers
with standard system calls, such as {\small \texttt{open}} and {\small\texttt{ioctl}}.
They perform coarse-grained computations while exchanging large data sets
with the memory hierarchy.
\figurename~\ref{fig:accelerator-struct} shows the structure and interface
common to all ESP accelerators.
The interface channels allow the accelerator to
(1) communicate with the CPU via memory-mapped registers ({\small \texttt{conf\_info}}),
(2) program the DMA controller or interact with other accelerators ({\small \texttt{load\_ctrl}} and {\small \texttt{store\_ctrl}}),
(3) exchange data with the memory hierarchy or other accelerators,
({\small\texttt{load\_chnl}} and {\small\texttt{store\_chnl}}),
and (4) notify its completion back to the software
application ({\small\texttt{acc\_done}}).

These channels are implemented with latency-insensitive communication
primitives, which HLS tools commonly provide as libraries
(e.g. Mentor MatchLib Connections~\cite{khailany_dac18}, Cadence Flex
Channels~\cite{meredith2008high}, Xilinx {\small\texttt{ap\_fifo}}).
These primitives preserve functional correctness in the presence of latency
variation both in the computation within the accelerator and in the
communication across the NoC~\cite{carloni_pieee15}.
This is obtained by adding \emph{valid} and \emph{ready} signals to the channels.
The valid signal indicates that the value of the data bundle is valid in the
current clock cycle, while the ready signal is de-asserted to apply backpressure.
The latency-insensitive nature of ESP accelerators allows designers to fully
exploit the ability of HLS to produce many alternative RTL implementations,
which are not strictly equivalent from an RTL viewpoint
(i.e., they do not produce the same timed sequence of outputs
for any valid input sequence), but they are members of a latency-equivalent
design class~\cite{carloni_tcad01_lip}. Each member of this class can be
seamlessly replaced with another one, depending on performance and cost
targets~\cite{piccolboni_tecs17}.

The execution flow of an ESP accelerator consists of four phases,
\emph{configure}, \emph{load}, \emph{compute}, and \emph{store}, as shown in
\figurename~\ref{fig:accelerator-struct}. A software application configures,
checks, and starts the accelerator via memory-mapped registers.
During the load and store phases, the accelerator interacts with the DMA
controller,
interleaving data exchanges between the system and the accelerator's private
local memory (PLM) with computation. When the accelerator completes its task, an
interrupt resumes the software for further processing.

For better performance, the accelerator can have one or more parallel
computation components that interact with the PLM.
The organization of the PLM itself is typically customized for the given
accelerator, with multiple banks and ports.
For example, the designer can organize it as a circular or ping-pong buffers to
support the pipelining of computation and transfers with the external memory or
other accelerators.
Designers can leverage PLM generators~\cite{pilato_tcad17} to implement many
different memory subsystems, each optimized for a specific combination of HLS
knobs settings.

\textbf{Accelerator Behavior.}
The charts of \figurename~\ref{fig:accelerator-comm} show the behavior of two
concurrent ESP accelerators ({\small \texttt{ACC0}} and {\small \texttt{ACC1}})
in three possible scenarios.
The two accelerators work in a producer-consumer fashion: {\small\texttt{ACC0}}
generates data that {\small\texttt{ACC1}} takes as inputs.
The accelerators are executed two times and concurrently; the consumer starts as
soon as the data is ready; finally, both the accelerators perform burst of
load and store DMA transactions, in red and brown respectively.
The completion of the configuration phase and interrupt request
({\small\texttt{acc\_done}})
are marked with {\small\texttt{CFG}} and {\small\texttt{IRQ}}, respectively.

In the top scenario, the two accelerators communicate via external memory.
First, the producer {\small\texttt{ACC0}} runs and stores the resulting data in main
memory. Upon completion of the producer, the consumer {\small\texttt{ACC1}} starts and
accesses the data in main memory; concurrently, the producer {\small\texttt{ACC0}} can
run a second time. The data exchange happens through memory at the granularity
of the whole accelerator data set.
This scenario is a \emph{virtual pipeline of ESP accelerators through memory}.
Ping-pong buffering on the PLM for both load and store phases allows the overlap
of computation and communication. In addition, load and store phases are allowed
to overlap. This is only possible by assuming to have dedicated memory channels
for each accelerator (e.g. two ESP memory tiles).
As long as the NoC and memory bandwidth are not saturated, the performance
overhead is limited only to the driver run time and the interrupt handling
procedures. We consider this scenario ideal.

In complex SoCs, it is reasonable to expect resource contention and delays with
the main memory. This can potentially limit the latency and throughput of the
accelerators, as shown in the middle scenario of
\figurename~\ref{fig:accelerator-comm}, where some of the DMA transactions get
delayed for both the producer and consumer accelerators.
The ESP library and API allows designers to replace the described software pipeline, with an
actual pipeline of accelerators, based on \emph{point-to-point communication
(P2P) over the NoC}.
The communication method does not need to be chosen at design time; instead,
special configuration registers are used to overwrite the default DMA behavior.
Beside relieving memory contention, P2P communication can actually improve
latency and throughput of communicating accelerators, as shown in the bottom
scenario of \figurename~\ref{fig:accelerator-comm}.
Here, each output transaction of the producer {\small\texttt{ACC0}} is matched to an
input transaction of the consumer {\small\texttt{ACC1}} (in green). Differently from the
previous scenarios, the data exchange via P2P happens at a smaller granularity:
a single store transaction of the producer {\small\texttt{ACC0}} is a valid input for
the consumer {\small\texttt{ACC1}}. A designer should take into account this assumption
when designing accelerators for a specific task.

\begin{figure*}[t!]
  \centering
  \includegraphics[width=1.0\textwidth]{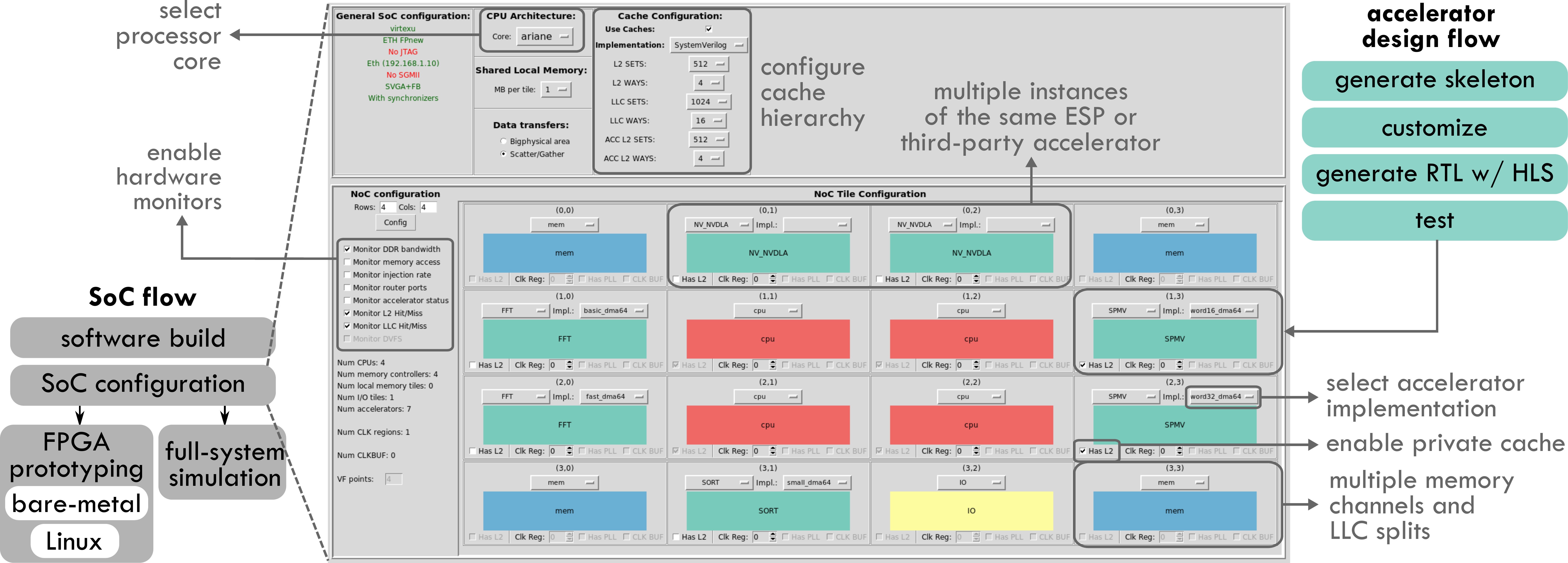}
    \vspace{-0.2in}
  \caption{Overview of the accelerator and SoC design flows with an example of
    SoC design configuration on the ESP GUI.}
  \label{fig:flows}
\end{figure*}

\textbf{Accelerator Templates and Code Generator.}
ESP provides the designers with a set of accelerator templates for each of the
HLS-based design flows. These templates leverage concepts of object-oriented
programming and class inheritance to simplify the design of the accelerators
in C/C++ or SystemC and enforce the interface and structure previously
described. They also implicitly address the differences existing among the
various HLS tools and input specification languages. For example, the
latency-insensitive primitives, which come with the different vendors, may have
slightly different APIs, e.g.
{\small\texttt{Put()}}/{\small\texttt{Get()}} vs.
{\small\texttt{Read()}}/{\small\texttt{Write()}},
or timing behavior. With some HLS tools, the designer has to specify some extra
{\small\texttt{wait()}} statements in SystemC to generate the correct RTL code.
In the case of C/C++ designs a combination of HLS directives and coding style
must be followed to ensure that extra memories are not inadvertently inferred
and the phases are correctly synchronized.

Next to templates, ESP provides a further aid for the accelerator design:
an interactive environment that generates a fully-working and HLS-ready
accelerator {\em skeleton} from a set of parameters passed by the designer.
The skeleton comes with a unit testbench, synthesis and simulation scripts, a
bare-metal driver, a Linux driver, and a sample test application.
This is the first step of the accelerator design flow, as shown on the top-right
of \figurename~\ref{fig:flows}.
The skeleton is a basic specification that uses the templates and contains
placeholder for manual accelerator-specific customizations.
The parameters passed by the designers include: unique name and ID, desired HLS
tool flow, a list of application-specific configuration registers, bit-width of
the data tokens, size of the data set and number of batches of data sets to be
executed without interrupting the CPU.
Next to application-specific information, designers can choose architectural
parameters that set the minimum required size of the PLM and the maximum
memory footprint of the application that invokes the accelerator. These
parameters have effect on the generated accelerator skeleton, device-driver,
test application, and on the configuration parameters for the ESP socket that
will host the accelerator.

Starting from the automatically generated skeleton, designers must customize the
accelerator computation phase, leveraging the software implementation of the
target computation kernel as a reference.
In addition, they are responsible for customizing the input generation and
output validation functions in the unit testbench and in the bare-metal and
Linux test applications. Finally, in case of complex data access patterns,
they may also need to extend the communication part of the accelerator and
define a more complex structure for the PLM.
The ESP release offers a set of online tutorials that describe these steps in
details with simple examples, which demonstrate how the first version of a new
accelerator can be designed, integrated and tested on FPGA in a few hours~\cite{esp}.

The domain specific flow for embedded machine learning is fully automated~\cite{giri_date20}.
The accelerator and the related software drivers and application are generated
in their entirety from the neural-network model. ESP automatically generates
also the accelerator tile socket and a wrapper for the accelerator logic.

\subsection{Third-Party Accelerator Integration} \label{sec:tp-acc-integration}
For existing accelerators, ESP provides a third-party accelerator integration
flow (TPF). The TPF skips all the steps necessary to design a new
accelerator and goes directly to SoC integration.
The designer must provide some information about the existing IP block and a
simple wrapper to connect the wires of the accelerator's interface to the ESP socket.
Specifically, the designer must fill in a short XML file with a unique
accelerator name and ID, the list and polarity of the reset signals, the list of
clock signals, an optional prefix for the AXI master interface in the wrapper, the
user-defined width of AXI optional control signals and the type of interrupt
request (i.e., level or edge sensitive).
In addition, the TPF requires the list of RTL source files, including Verilog,
SystemVerilog, VHDL and VHDL packages, a custom Makefile to compile the
third-party software and device drivers, and the list of executable files,
libraries and other binary objects needed to control the accelerator.

Currently, ESP provides adapters for AXI master (32 and 64 bits),
AHB master (32 bits) and AXI-Lite or APB slave (32 bits).
As long as the target accelerator is compliant with these standard bus
protocols, the Verilog top-level wrapper consists of a simple wire assignment to
expose bus ports to the ESP socket and connect any non-standard input port of
the third-party accelerator (e.g. disable test mode), if present.
After these simple manual steps, ESP takes care of the whole integration
automatically. We used the TPF to integrate the NVDLA~\cite{nvdla}.
An online tutorial in the ESP release demonstrates the design of a complete SoC
with multiple NVDLA tiles, multiple memory tiles and the Ariane RISC-V
processor. This system can run up to four concurrent machine-learning tasks
using the original NVDLA software stack\footnote{A minor patch was required to
  run multiple NVDLAs in a Linux environment.}~\cite{giri_carrv20}.

\subsection{SoC Flow} \label{sec:soc-flow}
The center and the left portion of \figurename~\ref{fig:flows} illustrate the
agile SoC development enabled by ESP.
Both the ESP and third-party accelerator flows contribute to the pool of
IP components that can be selected to build an SoC instance.
The ESP GUI guides the designers through an interactive SoC design flow that
allows them to: choose the number, types and positions of tiles, select
the desired Pareto-optimal design point from the HLS flows for each accelerator,
select the desired processor core among those available, determine the cache
hierarchy configuration, select the clock domains for each tile, and enable
the desired system monitors.
The GUI writes a configuration file that the ESP build flow can include to
generate RTL sockets, the system memory mapping, NoC routing tables,
the device tree for the target processor architecture, software header files, and
configuration parameters for the proxy components.

A single {\small \texttt{make}} target is sufficient to generate the bitstream for
one of the supported Xilinx evaluation boards (VCU128, VCU118 and VC707) and
proFPGA prototyping FPGA modules (XCVU440 and XC7V2000T).
Another single {\small \texttt{make}} target compiles Linux and creates a
default root file system that includes accelerators' drivers and test
applications, together with all necessary initialization scripts to load the ESP
library and memory allocator. If properly listed during the TPF, the software
stack for the third-party accelerators is loaded into the Linux image as well.
When the FPGA implementation is ready, users can load the
boot loader onto the ESP boot memory and the Linux image onto the
external DRAM with the \emph{ESP Link} application and the companion module on
the auxiliary tile.
Next
\emph{ESP Link} sends a soft reset to the processor cores, thus starting the
execution from the boot loader. Users can monitor the boot process via UART, or
log in with SSH after Linux boot completes. The online tutorials explain how to
properly wire the FPGA boards to a simple home router to ensure connectivity.

In addition to FPGA prototyping, designers can run full-system RTL simulations of
a bare-metal program. If monitoring the FPGA with the UART serial interface,
they can run bare-metal applications on FPGA as well.
The development of bare-metal and Linux applications for an ESP SoC is facilitated
by the ESP software stack described in Section~\ref{sec:software}.
The ESP release offers several examples.

The agile ESP flow allowed us to rapidly prototype many complex SoCs on FPGA,
including:

\begin{itemize}
\item An SoC with 12 computer vision accelerators, with as many dynamic frequency
  scaling (DFS) domains~\cite{mantovani_dac16}.
\item A multi-core SoC booting Linux SMP with tens of accelerators, multiple
  DRAM controllers, and dynamically reconfigurable cache coherence
  models~\cite{giri_aspdac19}.
\item A RISC-V based SoCs where deep learning applications running on top of
  Linux invoke loosely-coupled accelerators designed with multiple ESP
  accelerator design flows~\cite{giri_date20}.
\item A RISC-V based SoCs with multiple instances of the {\sc NVDLA}
  controlled by the RISC-V Ariane processor~\cite{giri_carrv20}.
\end{itemize}

\section{Related Work}
\label{sec:related}
The OSH movement is supported by multiple SoC design platforms, many based on the
RISC-V open-standard ISA~\cite{asanovic_case14,greengard20}.
The {\em Rocket Chip Generator} is an OSH project that leverages the Chisel RTL
language to construct SoCs with multiple RISC-V cores connected through a
coherent TileLink bus~\cite{lee_ieeemicro16}.
The {\em Chipyard} framework inherits Rocket Chip's Chisel-based parameterized
hardware generator methodology and also allows the integration of IP blocks
written in other RTL languages, via a Chisel wrapper, as well as
domain-specific accelerators ~\cite{amid_ieeemicro20}.
{\em Celerity} used the custom co-processor interface {\em RoCC} of the Rocket
chip to integrate five Rocket cores with an array of 496 simpler RISC-V cores
and a binarized neural network (BNN) accelerator, which was designed with HLS,
into a 385-million transistor SoC~\cite{davidson18}.
{\em HERO} is an FPGA-based research platform that allows the integration of a
standard host multicore processor with programmable manycore accelerators
composed of clusters of RISC-V cores based on the PULP
platform~\cite{kurth_carrv17,pulp,pulp_paper}.
{\em OpenPiton} was the first open-source SMP Linux-booting RISC-V multicore
processor~\cite{balkind_carrv19}.
It supports the research of heterogeneous ISAs and provides a coherence protocol
that extends across multiple chips~\cite{balkind_ieeemicro20,openpiton}.
{\em Blackparrot} is a multicore RISC-V architecture that offers some support for
the integration of loosely-coupled accelerators~\cite{petrisko_ieeemicro20};
currently, it provides two of the four cache-coherence options supported by ESP:
fully-coherent and non-coherent.

While most of these platforms are built with a processor-centric perspective,
ESP promotes a system-centric perspective with a scalable NoC-based architecture
and a strong focus on the integration of heterogeneous components, including
particularly loosely-coupled accelerators.
Another feature distinguishing ESP from the other open-source SoC platforms is
the flexible system-level design methodology that embraces a variety of
specification languages and synthesis flows, while promoting the use of HLS to
facilitate the design and integration of accelerators.

\section{Conclusions}
\label{sec:conclusions}

In summary, with ESP we aim at contributing to the open-source movement
by supporting the realization of more scalable architectures for SoCs that
integrate more heterogeneous components, thanks to a more flexible design
methodology that accommodates different specification languages and design flows.
Conceived as a heterogeneous integration platform and tested through years of
teaching at Columbia University, ESP is naturally suited to foster collaborative
engineering of SoCs across the OSH community.

\begin{acks}
{
  Over the years, the ESP project has been supported in part by
  DARPA (C\#: HR001113C0003 and HR001118C0122),
  the ARO (G\#: W911NF-19-1-0476),
  the NSF (A\#: 1219001),
  and C-FAR (C\#:2013-MA-2384), an SRC STARnet center.
The views and conclusions contained in this document are those of the authors
and should not be interpreted as representing the official policies, either
expressed or implied, of the Army Research Office, the Department of Defense or the U.S. Government.
The U.S. Government is authorized to reproduce and distribute reprints for
Government purposes notwithstanding any copyright notation herein.
}
\end{acks}

{
\balance
\bibliographystyle{ACM-Reference-Format}
\bibliography{paper}


\begin{thebibliography}{51}


\ifx \showCODEN    \undefined \def \showCODEN     #1{\unskip}     \fi
\ifx \showDOI      \undefined \def \showDOI       #1{#1}\fi
\ifx \showISBNx    \undefined \def \showISBNx     #1{\unskip}     \fi
\ifx \showISBNxiii \undefined \def \showISBNxiii  #1{\unskip}     \fi
\ifx \showISSN     \undefined \def \showISSN      #1{\unskip}     \fi
\ifx \showLCCN     \undefined \def \showLCCN      #1{\unskip}     \fi
\ifx \shownote     \undefined \def \shownote      #1{#1}          \fi
\ifx \showarticletitle \undefined \def \showarticletitle #1{#1}   \fi
\ifx \showURL      \undefined \def \showURL       {\relax}        \fi
\providecommand\bibfield[2]{#2}
\providecommand\bibinfo[2]{#2}
\providecommand\natexlab[1]{#1}
\providecommand\showeprint[2][]{arXiv:#2}

\bibitem[\protect\citeauthoryear{??}{ari}{[n.d.]}]%
        {ariane}
\bibinfo{title}{{Ariane}}.
\newblock \bibinfo{howpublished}{{\scriptsize
  \url{https://github.com/pulp-platform/ariane}}}.
\newblock


\bibitem[\protect\citeauthoryear{??}{hls}{[n.d.]}]%
        {hls4ml}
\bibinfo{title}{HLS4ML}.
\newblock \bibinfo{howpublished}{{\scriptsize
  \url{https://fastmachinelearning.org/hls4ml/}}}.
\newblock


\bibitem[\protect\citeauthoryear{??}{nvd}{[n.d.]}]%
        {nvdla}
\bibinfo{title}{{{NVIDIA} Deep Learning Accelerator (NVDLA)}}.
\newblock \bibinfo{howpublished}{{\scriptsize \url{www.nvdla.org}}}.
\newblock


\bibitem[\protect\citeauthoryear{??}{pul}{[n.d.]}]%
        {pulp}
\bibinfo{title}{{PULP}}.
\newblock \bibinfo{howpublished}{{\scriptsize
  \url{https://pulp-platform.org/}}}.
\newblock


\bibitem[\protect\citeauthoryear{{Amid}, {Biancolin}, {Gonzalez}, {Grubb},
  {Karandikar}, {Liew}, {Magyar}, {Mao}, {Ou}, {Pemberton}, {Rigge}, {Schmidt},
  {Wright}, {Zhao}, {Shao}, {Asanovic}, and {Nikolic}}{{Amid}
  et~al\mbox{.}}{2020}]%
        {amid_ieeemicro20}
\bibfield{author}{\bibinfo{person}{A. {Amid}}, \bibinfo{person}{D.
  {Biancolin}}, \bibinfo{person}{A. {Gonzalez}}, \bibinfo{person}{D. {Grubb}},
  \bibinfo{person}{S. {Karandikar}}, \bibinfo{person}{H. {Liew}},
  \bibinfo{person}{A. {Magyar}}, \bibinfo{person}{H. {Mao}},
  \bibinfo{person}{A. {Ou}}, \bibinfo{person}{N. {Pemberton}},
  \bibinfo{person}{P. {Rigge}}, \bibinfo{person}{C. {Schmidt}},
  \bibinfo{person}{J. {Wright}}, \bibinfo{person}{J. {Zhao}},
  \bibinfo{person}{Y.~S. {Shao}}, \bibinfo{person}{K. {Asanovic}}, {and}
  \bibinfo{person}{B. {Nikolic}}.} \bibinfo{year}{2020}\natexlab{}.
\newblock \showarticletitle{{Chipyard: Integrated Design, Simulation, and
  Implementation Framework for Custom SoCs}}.
\newblock \bibinfo{journal}{\emph{IEEE Micro}} \bibinfo{volume}{40},
  \bibinfo{number}{4} (\bibinfo{year}{2020}), \bibinfo{pages}{10--21}.
\newblock


\bibitem[\protect\citeauthoryear{Asanovic and Patterson}{Asanovic and
  Patterson}{2014}]%
        {asanovic_case14}
\bibfield{author}{\bibinfo{person}{K. Asanovic} {and} \bibinfo{person}{D.
  Patterson}.} \bibinfo{year}{2014}\natexlab{}.
\newblock \showarticletitle{{The Case for Open Instruction Sets}}.
\newblock \bibinfo{journal}{\emph{Microprocessor Report}} (\bibinfo{date}{Aug.}
  \bibinfo{year}{2014}).
\newblock


\bibitem[\protect\citeauthoryear{{Balkind}, {Chang}, {Jackson},
  {Tziantzioulis}, {Li}, {Gao}, {Lavrov}, {Chirkov}, {Tu}, {Shahrad}, and
  {Wentzlaff}}{{Balkind} et~al\mbox{.}}{2020}]%
        {balkind_ieeemicro20}
\bibfield{author}{\bibinfo{person}{J. {Balkind}}, \bibinfo{person}{T. {Chang}},
  \bibinfo{person}{P.~J. {Jackson}}, \bibinfo{person}{G. {Tziantzioulis}},
  \bibinfo{person}{A. {Li}}, \bibinfo{person}{F. {Gao}}, \bibinfo{person}{A.
  {Lavrov}}, \bibinfo{person}{G. {Chirkov}}, \bibinfo{person}{J. {Tu}},
  \bibinfo{person}{M. {Shahrad}}, {and} \bibinfo{person}{D. {Wentzlaff}}.}
  \bibinfo{year}{2020}\natexlab{}.
\newblock \showarticletitle{OpenPiton at 5: A Nexus for Open and Agile Hardware
  Design}.
\newblock \bibinfo{journal}{\emph{IEEE Micro}} \bibinfo{volume}{40},
  \bibinfo{number}{4} (\bibinfo{year}{2020}), \bibinfo{pages}{22--31}.
\newblock


\bibitem[\protect\citeauthoryear{Balkind, Lim, Gao, Tu, Wentzlaff, Schaffner,
  Zaruba, and Benini}{Balkind et~al\mbox{.}}{2019}]%
        {balkind_carrv19}
\bibfield{author}{\bibinfo{person}{J. Balkind}, \bibinfo{person}{K. Lim},
  \bibinfo{person}{F. Gao}, \bibinfo{person}{J. Tu}, \bibinfo{person}{D.
  Wentzlaff}, \bibinfo{person}{M. Schaffner}, \bibinfo{person}{F. Zaruba},
  {and} \bibinfo{person}{L. Benini}.} \bibinfo{year}{2019}\natexlab{}.
\newblock \showarticletitle{{OpenPiton+ Ariane: The First Open-Source, SMP
  Linux-booting RISC-V System Scaling From One to Many Cores}}. In
  \bibinfo{booktitle}{\emph{Workshop on Computer Architecture Research with
  RISC-V (CARRV)}}. \bibinfo{pages}{1--6}.
\newblock


\bibitem[\protect\citeauthoryear{Borkar and Chen}{Borkar and Chen}{2011}]%
        {borkar11}
\bibfield{author}{\bibinfo{person}{S. Borkar} {and} \bibinfo{person}{A. Chen}.}
  \bibinfo{year}{2011}\natexlab{}.
\newblock \showarticletitle{The Future of Microprocessors}.
\newblock \bibinfo{journal}{\emph{Communication of the ACM}}
  \bibinfo{volume}{54} (\bibinfo{date}{May} \bibinfo{year}{2011}),
  \bibinfo{pages}{67--77}.
\newblock
Issue 5.


\bibitem[\protect\citeauthoryear{Carloni}{Carloni}{2015}]%
        {carloni_pieee15}
\bibfield{author}{\bibinfo{person}{L.~P. Carloni}.}
  \bibinfo{year}{2015}\natexlab{}.
\newblock \showarticletitle{{From Latency-Insensitive Design to
  Communication-Based System-Level Design}}.
\newblock \bibinfo{journal}{\emph{{Proceedings of the IEEE}}}
  \bibinfo{volume}{103}, \bibinfo{number}{11} (\bibinfo{date}{Nov.}
  \bibinfo{year}{2015}), \bibinfo{pages}{2133--2151}.
\newblock


\bibitem[\protect\citeauthoryear{Carloni}{Carloni}{2016}]%
        {carloni_dac16}
\bibfield{author}{\bibinfo{person}{L.~P. Carloni}.}
  \bibinfo{year}{2016}\natexlab{}.
\newblock \showarticletitle{{The Case for {Embedded Scalable Platforms}}}. In
  \bibinfo{booktitle}{\emph{Proc. of the Design Automation Conference (DAC)}}.
  \bibinfo{pages}{17:1--17:6}.
\newblock


\bibitem[\protect\citeauthoryear{Carloni, Cota, {Di Guglielmo}, Giri, Kwon,
  Mantovani, Piccolboni, and Petracca}{Carloni et~al\mbox{.}}{2019}]%
        {carloni_wcae19}
\bibfield{author}{\bibinfo{person}{L.~P. Carloni}, \bibinfo{person}{E.~G.
  Cota}, \bibinfo{person}{G. {Di Guglielmo}}, \bibinfo{person}{D. Giri},
  \bibinfo{person}{J. Kwon}, \bibinfo{person}{P. Mantovani},
  \bibinfo{person}{L. Piccolboni}, {and} \bibinfo{person}{M. Petracca}.}
  \bibinfo{year}{2019}\natexlab{}.
\newblock \showarticletitle{{Teaching Heterogeneous Computing with System-Level
  Design Methods}}. In \bibinfo{booktitle}{\emph{Workshop on Computer
  Architecture Education (WCAE)}}. \bibinfo{pages}{1--8}.
\newblock


\bibitem[\protect\citeauthoryear{Carloni, McMillan, Saldahna, and
  Sangiovanni-Vincentelli}{Carloni et~al\mbox{.}}{1999}]%
        {carloni_iccad99}
\bibfield{author}{\bibinfo{person}{L.~P. Carloni}, \bibinfo{person}{K.~L.
  McMillan}, \bibinfo{person}{A. Saldahna}, {and} \bibinfo{person}{A.~L.
  Sangiovanni-Vincentelli}.} \bibinfo{year}{1999}\natexlab{}.
\newblock \showarticletitle{A Methodology for {``}Correct-by-Construction{"}
  Latency Insensitive Design}. In \bibinfo{booktitle}{\emph{Proc. of the
  International Conference on Computer-Aided Design}}.
  \bibinfo{pages}{309--315}.
\newblock


\bibitem[\protect\citeauthoryear{Carloni, McMillan, and
  Sangiovanni-Vincentelli}{Carloni et~al\mbox{.}}{2001}]%
        {carloni_tcad01_lip}
\bibfield{author}{\bibinfo{person}{L.~P. Carloni}, \bibinfo{person}{K.~L.
  McMillan}, {and} \bibinfo{person}{A.~L. Sangiovanni-Vincentelli}.}
  \bibinfo{year}{2001}\natexlab{}.
\newblock \showarticletitle{Theory of Latency-Insensitive Design}.
\newblock \bibinfo{journal}{\emph{IEEE Transactions on CAD of Integrated
  Circuits and Systems}} \bibinfo{volume}{20}, \bibinfo{number}{9}
  (\bibinfo{date}{Sept.} \bibinfo{year}{2001}), \bibinfo{pages}{1059--1076}.
\newblock


\bibitem[\protect\citeauthoryear{{Caulfield}, {Chung}, {Putnam}, {Angepat},
  {Fowers}, {Haselman}, {Heil}, {Humphrey}, {Kaur}, {Kim}, {Lo}, {Massengill},
  {Ovtcharov}, {Papamichael}, {Woods}, {Lanka}, {Chiou}, and
  {Burger}}{{Caulfield} et~al\mbox{.}}{2016}]%
        {caulfield2016}
\bibfield{author}{\bibinfo{person}{A.~M. {Caulfield}}, \bibinfo{person}{E.~S.
  {Chung}}, \bibinfo{person}{A. {Putnam}}, \bibinfo{person}{H. {Angepat}},
  \bibinfo{person}{J. {Fowers}}, \bibinfo{person}{M. {Haselman}},
  \bibinfo{person}{S. {Heil}}, \bibinfo{person}{M. {Humphrey}},
  \bibinfo{person}{P. {Kaur}}, \bibinfo{person}{J. {Kim}}, \bibinfo{person}{D.
  {Lo}}, \bibinfo{person}{T. {Massengill}}, \bibinfo{person}{K. {Ovtcharov}},
  \bibinfo{person}{M. {Papamichael}}, \bibinfo{person}{L. {Woods}},
  \bibinfo{person}{S. {Lanka}}, \bibinfo{person}{D. {Chiou}}, {and}
  \bibinfo{person}{D. {Burger}}.} \bibinfo{year}{2016}\natexlab{}.
\newblock \showarticletitle{A Cloud-Scale Acceleration Architecture}. In
  \bibinfo{booktitle}{\emph{Proc. of the IEEE/ACM International Symposium on
  Microarchitecture (MICRO)}}. \bibinfo{pages}{1--13}.
\newblock


\bibitem[\protect\citeauthoryear{{Chishiro}, {Suito}, {Ito}, {Maeda}, {Azumi},
  {Funaoka}, and {Kato}}{{Chishiro} et~al\mbox{.}}{2019}]%
        {chishiro_ICESS19}
\bibfield{author}{\bibinfo{person}{H. {Chishiro}}, \bibinfo{person}{K.
  {Suito}}, \bibinfo{person}{T. {Ito}}, \bibinfo{person}{S. {Maeda}},
  \bibinfo{person}{T. {Azumi}}, \bibinfo{person}{K. {Funaoka}}, {and}
  \bibinfo{person}{S. {Kato}}.} \bibinfo{year}{2019}\natexlab{}.
\newblock \showarticletitle{{Towards Heterogeneous Computing Platforms for
  Autonomous Driving}}. In \bibinfo{booktitle}{\emph{{Proc. of the
  International Conference on Embedded Software and Systems (ICESS)}}}.
\newblock


\bibitem[\protect\citeauthoryear{{Cobham Gaisler}}{{Cobham Gaisler}}{[n.d.]}]%
        {leon3}
\bibfield{author}{\bibinfo{person}{{Cobham Gaisler}}.}
\newblock \bibinfo{title}{LEON3}.
\newblock \bibinfo{howpublished}{{\scriptsize
  \url{www.gaisler.com/index.php/products/processors/leon3}}}.
\newblock


\bibitem[\protect\citeauthoryear{{Columbia SLD Group}}{{Columbia SLD
  Group}}{2019}]%
        {esp}
\bibfield{author}{\bibinfo{person}{{Columbia SLD Group}}.}
\newblock \bibinfo{title}{{ESP Release}}.
\newblock \bibinfo{howpublished}{{\scriptsize \url{www.esp.cs.columbia.edu}}}.
\newblock


\bibitem[\protect\citeauthoryear{{Cota}, {Mantovani}, {Di Guglielmo}, and
  {Carloni}}{{Cota} et~al\mbox{.}}{2015}]%
        {cota_dac15}
\bibfield{author}{\bibinfo{person}{E.~G. {Cota}}, \bibinfo{person}{P.
  {Mantovani}}, \bibinfo{person}{G. {Di Guglielmo}}, {and}
  \bibinfo{person}{L.~P. {Carloni}}.} \bibinfo{year}{2015}\natexlab{}.
\newblock \showarticletitle{{An Analysis of Accelerator Coupling in
  Heterogeneous Architectures}}. In \bibinfo{booktitle}{\emph{Proc. of the
  Design Automation Conference (DAC)}}. \bibinfo{pages}{202:1--202:6}.
\newblock


\bibitem[\protect\citeauthoryear{Dally, Turakhia, and Han}{Dally
  et~al\mbox{.}}{2020}]%
        {dally2020}
\bibfield{author}{\bibinfo{person}{W. Dally}, \bibinfo{person}{Y. Turakhia},
  {and} \bibinfo{person}{S. Han}.} \bibinfo{year}{2020}\natexlab{}.
\newblock \showarticletitle{Domain-Specific Hardware Accelerators}.
\newblock \bibinfo{journal}{\emph{Communication of the ACM}}
  \bibinfo{volume}{63}, \bibinfo{number}{7} (\bibinfo{date}{June}
  \bibinfo{year}{2020}), \bibinfo{pages}{48--57}.
\newblock


\bibitem[\protect\citeauthoryear{{Davidson}, {Xie}, {Torng}, {Al-Hawai},
  {Rovinski}, {Ajayi}, {Vega}, {Zhao}, {Zhao}, {Dai}, {Amarnath}, {Veluri},
  {Gao}, {Rao}, {Liu}, {Gupta}, {Zhang}, {Dreslinski}, {Batten}, and
  {Taylor}}{{Davidson} et~al\mbox{.}}{2018}]%
        {davidson18}
\bibfield{author}{\bibinfo{person}{S. {Davidson}}, \bibinfo{person}{S. {Xie}},
  \bibinfo{person}{C. {Torng}}, \bibinfo{person}{K. {Al-Hawai}},
  \bibinfo{person}{A. {Rovinski}}, \bibinfo{person}{T. {Ajayi}},
  \bibinfo{person}{L. {Vega}}, \bibinfo{person}{C. {Zhao}}, \bibinfo{person}{R.
  {Zhao}}, \bibinfo{person}{S. {Dai}}, \bibinfo{person}{A. {Amarnath}},
  \bibinfo{person}{B. {Veluri}}, \bibinfo{person}{P. {Gao}},
  \bibinfo{person}{A. {Rao}}, \bibinfo{person}{G. {Liu}},
  \bibinfo{person}{R.~K. {Gupta}}, \bibinfo{person}{Z. {Zhang}},
  \bibinfo{person}{R. {Dreslinski}}, \bibinfo{person}{C. {Batten}}, {and}
  \bibinfo{person}{M.~B. {Taylor}}.} \bibinfo{year}{2018}\natexlab{}.
\newblock \showarticletitle{{The {Celerity} Open-Source {511-Core RISC-V}
  Tiered Accelerator Fabric: Fast Architectures and Design Methodologies for
  Fast Chips}}.
\newblock \bibinfo{journal}{\emph{IEEE Micro}} \bibinfo{volume}{38},
  \bibinfo{number}{2} (\bibinfo{date}{Feb.} \bibinfo{year}{2018}),
  \bibinfo{pages}{30--41}.
\newblock


\bibitem[\protect\citeauthoryear{Duarte, Han, Harris, Jindariani, Kreinar,
  Kreis, Ngadiuba, Pierini, Rivera, Tran, and Wu}{Duarte et~al\mbox{.}}{2018}]%
        {duarte_2018}
\bibfield{author}{\bibinfo{person}{J. Duarte}, \bibinfo{person}{S. Han},
  \bibinfo{person}{P. Harris}, \bibinfo{person}{S. Jindariani},
  \bibinfo{person}{E. Kreinar}, \bibinfo{person}{B. Kreis}, \bibinfo{person}{J.
  Ngadiuba}, \bibinfo{person}{M. Pierini}, \bibinfo{person}{R. Rivera},
  \bibinfo{person}{N. Tran}, {and} \bibinfo{person}{Z. Wu}.}
  \bibinfo{year}{2018}\natexlab{}.
\newblock \showarticletitle{Fast inference of deep neural networks in {FPGAs}
  for particle physics}.
\newblock \bibinfo{journal}{\emph{Journal of Instrumentation}}
  \bibinfo{volume}{13}, \bibinfo{number}{07} (\bibinfo{date}{July}
  \bibinfo{year}{2018}), \bibinfo{pages}{P07027--P07027}.
\newblock


\bibitem[\protect\citeauthoryear{Gaisler}{Gaisler}{2004}]%
        {grlib}
\bibfield{author}{\bibinfo{person}{J. Gaisler}.}
  \bibinfo{year}{2004}\natexlab{}.
\newblock \showarticletitle{An Open-Source {{VHDL IP}} Library with Plug \&
  Play Configuration}.
\newblock \bibinfo{journal}{\emph{Building the Information Society}}
  (\bibinfo{year}{2004}).
\newblock


\bibitem[\protect\citeauthoryear{Giri, Chiu, Eichler, Mantovani, Chandramoorth,
  and Carloni}{Giri et~al\mbox{.}}{2020a}]%
        {giri_carrv20}
\bibfield{author}{\bibinfo{person}{D. Giri}, \bibinfo{person}{K.-L. Chiu},
  \bibinfo{person}{G. Eichler}, \bibinfo{person}{P. Mantovani},
  \bibinfo{person}{N. Chandramoorth}, {and} \bibinfo{person}{L.~P. Carloni}.}
  \bibinfo{year}{2020}\natexlab{a}.
\newblock \showarticletitle{{Ariane + NVDLA: Seamless Third-Party IP
  Integration with ESP}}. In \bibinfo{booktitle}{\emph{Workshop on Computer
  Architecture Research with RISC-V (CARRV)}}.
\newblock


\bibitem[\protect\citeauthoryear{Giri, Chiu, Guglielmo, Mantovani, and
  Carloni}{Giri et~al\mbox{.}}{2020b}]%
        {giri_date20}
\bibfield{author}{\bibinfo{person}{D. Giri}, \bibinfo{person}{K.-L. Chiu},
  \bibinfo{person}{G.~Di Guglielmo}, \bibinfo{person}{P. Mantovani}, {and}
  \bibinfo{person}{L.P. Carloni}.} \bibinfo{year}{2020}\natexlab{b}.
\newblock \showarticletitle{{ESP4ML: Platform-Based Design of Systems-on-Chip
  for Embedded Machine Learning}}. In \bibinfo{booktitle}{\emph{Proc. of the
  Conference on Design, Automation, and Test in Europe (DATE)}}.
  \bibinfo{pages}{1049--1054}.
\newblock


\bibitem[\protect\citeauthoryear{Giri, Mantovani, and Carloni}{Giri
  et~al\mbox{.}}{2018a}]%
        {giri_ieeemicro18}
\bibfield{author}{\bibinfo{person}{D. Giri}, \bibinfo{person}{P. Mantovani},
  {and} \bibinfo{person}{L.~P. Carloni}.} \bibinfo{year}{2018}\natexlab{a}.
\newblock \showarticletitle{{Accelerators {\&} Coherence: An {SoC}
  Perspective}}.
\newblock \bibinfo{journal}{\emph{IEEE Micro}} \bibinfo{volume}{38},
  \bibinfo{number}{6} (\bibinfo{date}{Nov.} \bibinfo{year}{2018}),
  \bibinfo{pages}{36--45}.
\newblock


\bibitem[\protect\citeauthoryear{Giri, Mantovani, and Carloni}{Giri
  et~al\mbox{.}}{2018b}]%
        {giri_nocs18}
\bibfield{author}{\bibinfo{person}{D. Giri}, \bibinfo{person}{P. Mantovani},
  {and} \bibinfo{person}{L.~P. Carloni}.} \bibinfo{year}{2018}\natexlab{b}.
\newblock \showarticletitle{{{NoC}-Based Support of Heterogeneous
  Cache-Coherence Models for Accelerators}}. In \bibinfo{booktitle}{\emph{Proc.
  of the International Symposium on Networks-on-Chip (NOCS)}}.
  \bibinfo{pages}{1:1--1:8}.
\newblock


\bibitem[\protect\citeauthoryear{Giri, Mantovani, and Carloni}{Giri
  et~al\mbox{.}}{2019}]%
        {giri_aspdac19}
\bibfield{author}{\bibinfo{person}{D. Giri}, \bibinfo{person}{P. Mantovani},
  {and} \bibinfo{person}{L.~P. Carloni}.} \bibinfo{year}{2019}\natexlab{}.
\newblock \showarticletitle{{Runtime Reconfigurable Memory Hierarchy in
  Embedded Scalable Platforms}}. In \bibinfo{booktitle}{\emph{Proc. of the Asia
  and South Pacific Design Automation Conference (ASPDAC)}}.
  \bibinfo{pages}{719--726}.
\newblock


\bibitem[\protect\citeauthoryear{Greengard}{Greengard}{2020}]%
        {greengard20}
\bibfield{author}{\bibinfo{person}{S. Greengard}.}
  \bibinfo{year}{2020}\natexlab{}.
\newblock \showarticletitle{Will {RISC-V} Revolutionize Computing?}
\newblock \bibinfo{journal}{\emph{Commun. ACM}} \bibinfo{volume}{63},
  \bibinfo{number}{5} (\bibinfo{date}{April} \bibinfo{year}{2020}),
  \bibinfo{pages}{30--32}.
\newblock


\bibitem[\protect\citeauthoryear{{Gupta}, {Nowatzki}, {Gangadhar}, and
  {Sankaralingam}}{{Gupta} et~al\mbox{.}}{2017}]%
        {gupta17}
\bibfield{author}{\bibinfo{person}{G. {Gupta}}, \bibinfo{person}{T.
  {Nowatzki}}, \bibinfo{person}{V. {Gangadhar}}, {and} \bibinfo{person}{K.
  {Sankaralingam}}.} \bibinfo{year}{2017}\natexlab{}.
\newblock \showarticletitle{{Kickstarting Semiconductor Innovation with Open
  Source Hardware}}.
\newblock \bibinfo{journal}{\emph{IEEE Computer}} \bibinfo{volume}{50},
  \bibinfo{number}{6} (\bibinfo{date}{June} \bibinfo{year}{2017}),
  \bibinfo{pages}{50--59}.
\newblock


\bibitem[\protect\citeauthoryear{Horowitz}{Horowitz}{2014}]%
        {Horowitz2014}
\bibfield{author}{\bibinfo{person}{M. Horowitz}.}
  \bibinfo{year}{2014}\natexlab{}.
\newblock \showarticletitle{Computing's Energy Problem (and What We Can Do
  About It)}. In \bibinfo{booktitle}{\emph{International Solid-State Circuits
  Conference (ISSCC)}}. \bibinfo{pages}{10--14}.
\newblock


\bibitem[\protect\citeauthoryear{Jouppi, Young, Patil, and Patterson}{Jouppi
  et~al\mbox{.}}{2018}]%
        {jouppi2018}
\bibfield{author}{\bibinfo{person}{N.~P. Jouppi}, \bibinfo{person}{C. Young},
  \bibinfo{person}{N. Patil}, {and} \bibinfo{person}{D. Patterson}.}
  \bibinfo{year}{2018}\natexlab{}.
\newblock \showarticletitle{{A Domain-Specific Architecture for Deep Neural
  Networks}}.
\newblock \bibinfo{journal}{\emph{Commun. ACM}} \bibinfo{volume}{61},
  \bibinfo{number}{9} (\bibinfo{date}{Aug.} \bibinfo{year}{2018}),
  \bibinfo{pages}{50--59}.
\newblock


\bibitem[\protect\citeauthoryear{{Khailany}, {Krimer}, {Venkatesan}, {Clemons},
  {Emer}, {Fojtik}, {Klinefelter}, {Pellauer}, {Pinckney}, {Shao}, {Srinath},
  {Torng}, {Xi}, {Zhang}, and {Zimmer}}{{Khailany} et~al\mbox{.}}{2018}]%
        {khailany_dac18}
\bibfield{author}{\bibinfo{person}{B. {Khailany}}, \bibinfo{person}{E.
  {Krimer}}, \bibinfo{person}{R. {Venkatesan}}, \bibinfo{person}{J. {Clemons}},
  \bibinfo{person}{J.~S. {Emer}}, \bibinfo{person}{M. {Fojtik}},
  \bibinfo{person}{A. {Klinefelter}}, \bibinfo{person}{M. {Pellauer}},
  \bibinfo{person}{N. {Pinckney}}, \bibinfo{person}{Y.~S. {Shao}},
  \bibinfo{person}{S. {Srinath}}, \bibinfo{person}{C. {Torng}},
  \bibinfo{person}{S.~L. {Xi}}, \bibinfo{person}{Y. {Zhang}}, {and}
  \bibinfo{person}{B. {Zimmer}}.} \bibinfo{year}{2018}\natexlab{}.
\newblock \showarticletitle{{A Modular Digital VLSI Flow for High-Productivity
  SoC Design}}. In \bibinfo{booktitle}{\emph{Proc. of the Design Automation
  Conference (DAC)}}. \bibinfo{pages}{1--6}.
\newblock


\bibitem[\protect\citeauthoryear{Kurth, Vogel, Capotondi, Marongui, and
  Benini}{Kurth et~al\mbox{.}}{2017}]%
        {kurth_carrv17}
\bibfield{author}{\bibinfo{person}{A. Kurth}, \bibinfo{person}{P. Vogel},
  \bibinfo{person}{A. Capotondi}, \bibinfo{person}{A. Marongui}, {and}
  \bibinfo{person}{L. Benini}.} \bibinfo{year}{2017}\natexlab{}.
\newblock \showarticletitle{{HERO: Heterogeneous Embedded Research Platform for
  Exploring RISC-V Manycore Accelerators on FPGA}}. In
  \bibinfo{booktitle}{\emph{Workshop on Computer Architecture Research with
  RISC-V (CARRV)}}. \bibinfo{pages}{1--7}.
\newblock


\bibitem[\protect\citeauthoryear{{Lee}, {Waterman}, {Cook}, {Zimmer}, {Keller},
  {Puggelli}, {Kwak}, {Jevtic}, {Bailey}, {Blagojevic}, {Chiu}, {Avizienis},
  {Richards}, {Bachrach}, {Patterson}, {Alon}, {Nikolic}, and {Asanovic}}{{Lee}
  et~al\mbox{.}}{2016}]%
        {lee_ieeemicro16}
\bibfield{author}{\bibinfo{person}{Y. {Lee}}, \bibinfo{person}{A. {Waterman}},
  \bibinfo{person}{H. {Cook}}, \bibinfo{person}{B. {Zimmer}},
  \bibinfo{person}{B. {Keller}}, \bibinfo{person}{A. {Puggelli}},
  \bibinfo{person}{J. {Kwak}}, \bibinfo{person}{R. {Jevtic}},
  \bibinfo{person}{S. {Bailey}}, \bibinfo{person}{M. {Blagojevic}},
  \bibinfo{person}{P. {Chiu}}, \bibinfo{person}{R. {Avizienis}},
  \bibinfo{person}{B. {Richards}}, \bibinfo{person}{J. {Bachrach}},
  \bibinfo{person}{D. {Patterson}}, \bibinfo{person}{E. {Alon}},
  \bibinfo{person}{B. {Nikolic}}, {and} \bibinfo{person}{K. {Asanovic}}.}
  \bibinfo{year}{2016}\natexlab{}.
\newblock \showarticletitle{{An Agile Approach to Building {RISC-V}
  Microprocessors}}.
\newblock \bibinfo{journal}{\emph{IEEE Micro}} \bibinfo{volume}{36},
  \bibinfo{number}{2} (\bibinfo{date}{Mar.-Apr.} \bibinfo{year}{2016}),
  \bibinfo{pages}{8--20}.
\newblock


\bibitem[\protect\citeauthoryear{Liu and Carloni}{Liu and Carloni}{2013}]%
        {liu_dac13}
\bibfield{author}{\bibinfo{person}{H-Y. Liu} {and} \bibinfo{person}{L.~P.
  Carloni}.} \bibinfo{year}{2013}\natexlab{}.
\newblock \showarticletitle{{On Learning-based Methods for Design-Space
  Exploration with High-Level Synthesis}}. In \bibinfo{booktitle}{\emph{Proc.
  of the Design Automation Conference (DAC)}}. \bibinfo{pages}{1--7}.
\newblock


\bibitem[\protect\citeauthoryear{Liu, Petracca, and Carloni}{Liu
  et~al\mbox{.}}{2012}]%
        {liu_date12}
\bibfield{author}{\bibinfo{person}{H-Y. Liu}, \bibinfo{person}{M. Petracca},
  {and} \bibinfo{person}{L.~P. Carloni}.} \bibinfo{year}{2012}\natexlab{}.
\newblock \showarticletitle{{Compositional System-Level Design Exploration with
  Planning of High-Level Synthesis}}. In \bibinfo{booktitle}{\emph{Proc. of the
  Conference on Design, Automation, and Test in Europe (DATE)}}.
  \bibinfo{pages}{641--646}.
\newblock


\bibitem[\protect\citeauthoryear{{M. Meredith}}{{M. Meredith}}{2008}]%
        {meredith2008high}
\bibfield{author}{\bibinfo{person}{{M. Meredith}}.}
  \bibinfo{year}{2008}\natexlab{}.
\newblock \showarticletitle{{High-level SystemC Synthesis with Forte's
  Cynthesizer}}.
\newblock In \bibinfo{booktitle}{\emph{{High-Level Synthesis}}}.
  \bibinfo{publisher}{Springer}, \bibinfo{pages}{75--97}.
\newblock


\bibitem[\protect\citeauthoryear{{Mantovani}, {Cota}, {Pilato}, {Di Guglielmo},
  and {Carloni}}{{Mantovani} et~al\mbox{.}}{2016a}]%
        {mantovani_cases16}
\bibfield{author}{\bibinfo{person}{P. {Mantovani}}, \bibinfo{person}{E.~G.
  {Cota}}, \bibinfo{person}{C. {Pilato}}, \bibinfo{person}{G. {Di Guglielmo}},
  {and} \bibinfo{person}{L.~P. {Carloni}}.} \bibinfo{year}{2016}\natexlab{a}.
\newblock \showarticletitle{Handling Large Data Sets for High-Performance
  Embedded Applications in Heterogeneous Systems-on-Chip}. In
  \bibinfo{booktitle}{\emph{Proc. of the Intl. Conference on Compilers,
  Architectures, and Synthesis of Embedded Systems (CASES)}}.
  \bibinfo{pages}{1--10}.
\newblock


\bibitem[\protect\citeauthoryear{{Mantovani}, {Cota}, {Tien}, {Pilato}, {Di
  Guglielmo}, {Shepard}, and {Carloni}}{{Mantovani} et~al\mbox{.}}{2016b}]%
        {mantovani_dac16}
\bibfield{author}{\bibinfo{person}{P. {Mantovani}}, \bibinfo{person}{E.~G.
  {Cota}}, \bibinfo{person}{K. {Tien}}, \bibinfo{person}{C. {Pilato}},
  \bibinfo{person}{G. {Di Guglielmo}}, \bibinfo{person}{K. {Shepard}}, {and}
  \bibinfo{person}{L.~P. {Carloni}}.} \bibinfo{year}{2016}\natexlab{b}.
\newblock \showarticletitle{An {FPGA}-Based Infrastructure for Fine-Grained
  {DVFS} Analysis in High-Performance Embedded Systems}. In
  \bibinfo{booktitle}{\emph{Proc. of the Design Automation Conference (DAC)}}.
  \bibinfo{pages}{157:1--157:6}.
\newblock


\bibitem[\protect\citeauthoryear{Mantovani, {Di Guglielmo}, and
  Carloni}{Mantovani et~al\mbox{.}}{2016}]%
        {mantovani_aspdac16}
\bibfield{author}{\bibinfo{person}{P. Mantovani}, \bibinfo{person}{G. {Di
  Guglielmo}}, {and} \bibinfo{person}{L.~P. Carloni}.}
  \bibinfo{year}{2016}\natexlab{}.
\newblock \showarticletitle{{High-level Synthesis of Accelerators in Embedded
  Scalable Platforms}}. In \bibinfo{booktitle}{\emph{Proc. of the Asia and
  South Pacific Design Automation Conference (ASPDAC)}}.
  \bibinfo{pages}{204--211}.
\newblock


\bibitem[\protect\citeauthoryear{{Nane}, {Sima}, {Pilato}, {Choi}, {Fort},
  {Canis}, {Chen}, {Hsiao}, {Brown}, {Ferrandi}, {Anderson}, and
  {Bertels}}{{Nane} et~al\mbox{.}}{2016}]%
        {nane16}
\bibfield{author}{\bibinfo{person}{R. {Nane}}, \bibinfo{person}{V. {Sima}},
  \bibinfo{person}{C. {Pilato}}, \bibinfo{person}{J. {Choi}},
  \bibinfo{person}{B. {Fort}}, \bibinfo{person}{A. {Canis}},
  \bibinfo{person}{Y.~T. {Chen}}, \bibinfo{person}{H. {Hsiao}},
  \bibinfo{person}{S. {Brown}}, \bibinfo{person}{F. {Ferrandi}},
  \bibinfo{person}{J. {Anderson}}, {and} \bibinfo{person}{K. {Bertels}}.}
  \bibinfo{year}{2016}\natexlab{}.
\newblock \showarticletitle{A Survey and Evaluation of FPGA High-Level
  Synthesis Tools}.
\newblock \bibinfo{journal}{\emph{IEEE Transactions on CAD of Integrated
  Circuits and Systems}} \bibinfo{volume}{35}, \bibinfo{number}{10}
  (\bibinfo{year}{2016}), \bibinfo{pages}{1591--1604}.
\newblock


\bibitem[\protect\citeauthoryear{{Petrisko}, {Gilani}, {Wyse}, {Jung},
  {Davidson}, {Gao}, {Zhao}, {Azad}, {Canakci}, {Veluri}, {Guarino}, {Joshi},
  {Oskin}, and {Taylor}}{{Petrisko} et~al\mbox{.}}{2020}]%
        {petrisko_ieeemicro20}
\bibfield{author}{\bibinfo{person}{D. {Petrisko}}, \bibinfo{person}{F.
  {Gilani}}, \bibinfo{person}{M. {Wyse}}, \bibinfo{person}{D.~C. {Jung}},
  \bibinfo{person}{S. {Davidson}}, \bibinfo{person}{P. {Gao}},
  \bibinfo{person}{C. {Zhao}}, \bibinfo{person}{Z. {Azad}}, \bibinfo{person}{S.
  {Canakci}}, \bibinfo{person}{B. {Veluri}}, \bibinfo{person}{T. {Guarino}},
  \bibinfo{person}{A. {Joshi}}, \bibinfo{person}{M. {Oskin}}, {and}
  \bibinfo{person}{M.~B. {Taylor}}.} \bibinfo{year}{2020}\natexlab{}.
\newblock \showarticletitle{BlackParrot: An Agile Open-Source RISC-V Multicore
  for Accelerator SoCs}.
\newblock \bibinfo{journal}{\emph{IEEE Micro}} \bibinfo{volume}{40},
  \bibinfo{number}{4} (\bibinfo{year}{2020}), \bibinfo{pages}{93--102}.
\newblock


\bibitem[\protect\citeauthoryear{Piccolboni, Mantovani, Guglielmo, and
  Carloni}{Piccolboni et~al\mbox{.}}{2017}]%
        {piccolboni_tecs17}
\bibfield{author}{\bibinfo{person}{L. Piccolboni}, \bibinfo{person}{P.
  Mantovani}, \bibinfo{person}{G.~Di Guglielmo}, {and} \bibinfo{person}{L.~P.
  Carloni}.} \bibinfo{year}{2017}\natexlab{}.
\newblock \showarticletitle{{COSMOS}: Coordination of High-Level Synthesis and
  Memory Optimization for Hardware Accelerators}.
\newblock \bibinfo{journal}{\emph{ACM Transactions on Embedded Computing
  Systems}} \bibinfo{volume}{16}, \bibinfo{number}{5s} (\bibinfo{date}{Sept.}
  \bibinfo{year}{2017}), \bibinfo{pages}{150:1--150:22}.
\newblock


\bibitem[\protect\citeauthoryear{{Pilato}, {Mantovani}, {Di Guglielmo}, and
  {Carloni}}{{Pilato} et~al\mbox{.}}{2017}]%
        {pilato_tcad17}
\bibfield{author}{\bibinfo{person}{C. {Pilato}}, \bibinfo{person}{P.
  {Mantovani}}, \bibinfo{person}{G. {Di Guglielmo}}, {and}
  \bibinfo{person}{L.~P. {Carloni}}.} \bibinfo{year}{2017}\natexlab{}.
\newblock \showarticletitle{{System-Level Optimization of Accelerator Local
  Memory for Heterogeneous Systems-on-Chip}}.
\newblock \bibinfo{journal}{\emph{IEEE Transactions on CAD of Integrated
  Circuits and Systems}} \bibinfo{volume}{36}, \bibinfo{number}{3}
  (\bibinfo{date}{March} \bibinfo{year}{2017}), \bibinfo{pages}{435--448}.
\newblock


\bibitem[\protect\citeauthoryear{{Princeton Parallel Group}}{{Princeton
  Parallel Group}}{[n.d.]}]%
        {openpiton}
\bibfield{author}{\bibinfo{person}{{Princeton Parallel Group}}.}
\newblock \bibinfo{title}{{OpenPiton}}.
\newblock \bibinfo{howpublished}{{\scriptsize
  \url{https://parallel.princeton.edu/openpiton/}}}.
\newblock


\bibitem[\protect\citeauthoryear{{Rossi}, {Loi}, {Conti}, {Tagliavini},
  {Pullini}, and {Marongiu}}{{Rossi} et~al\mbox{.}}{2014}]%
        {pulp_paper}
\bibfield{author}{\bibinfo{person}{D. {Rossi}}, \bibinfo{person}{I. {Loi}},
  \bibinfo{person}{F. {Conti}}, \bibinfo{person}{G. {Tagliavini}},
  \bibinfo{person}{A. {Pullini}}, {and} \bibinfo{person}{A. {Marongiu}}.}
  \bibinfo{year}{2014}\natexlab{}.
\newblock \showarticletitle{{Energy Efficient Parallel Computing on the PULP
  Platform with Support for OpenMP}}. In \bibinfo{booktitle}{\emph{Convention
  of Electrical Electronics Engineers in Israel (IEEEI)}}.
\newblock


\bibitem[\protect\citeauthoryear{Shao, Reagen, Wei, and Brooks}{Shao
  et~al\mbox{.}}{2015}]%
        {shao_ieeemicro_15}
\bibfield{author}{\bibinfo{person}{Y.~S. Shao}, \bibinfo{person}{B. Reagen},
  \bibinfo{person}{G. Wei}, {and} \bibinfo{person}{D. Brooks}.}
  \bibinfo{year}{2015}\natexlab{}.
\newblock \showarticletitle{{The Aladdin Approach to Accelerator Design and
  Modeling}}.
\newblock \bibinfo{journal}{\emph{IEEE Micro}} \bibinfo{volume}{35},
  \bibinfo{number}{3} (\bibinfo{date}{May-Jun} \bibinfo{year}{2015}),
  \bibinfo{pages}{58--70}.
\newblock


\bibitem[\protect\citeauthoryear{Yoon, Concer, and Carloni}{Yoon
  et~al\mbox{.}}{2013}]%
        {yoon_tcad13}
\bibfield{author}{\bibinfo{person}{Y.-J. Yoon}, \bibinfo{person}{N. Concer},
  {and} \bibinfo{person}{L.~P. Carloni}.} \bibinfo{year}{2013}\natexlab{}.
\newblock \showarticletitle{Virtual Channels and Multiple Physical Networks:
  Two Alternatives to Improve {NoC} Performance}.
\newblock \bibinfo{journal}{\emph{IEEE Transactions on CAD of Integrated
  Circuits and Systems}} \bibinfo{volume}{32}, \bibinfo{number}{12}
  (\bibinfo{date}{Dec.} \bibinfo{year}{2013}), \bibinfo{pages}{1906--1919}.
\newblock


\bibitem[\protect\citeauthoryear{Yoon, Mantovani, and Carloni}{Yoon
  et~al\mbox{.}}{2017}]%
        {yoon_nocs17}
\bibfield{author}{\bibinfo{person}{Y.-J. Yoon}, \bibinfo{person}{P. Mantovani},
  {and} \bibinfo{person}{L.~P. Carloni}.} \bibinfo{year}{2017}\natexlab{}.
\newblock \showarticletitle{{System-Level Design of Networks-on-Chip for
  Heterogeneous Systems-on-Chip}}. In \bibinfo{booktitle}{\emph{Proc. of the
  International Symposium on Networks-on-Chip (NOCS)}}. \bibinfo{pages}{1--6}.
\newblock


\bibitem[\protect\citeauthoryear{Zaruba and Benini}{Zaruba and Benini}{2019}]%
        {ariane_paper}
\bibfield{author}{\bibinfo{person}{F. Zaruba} {and} \bibinfo{person}{L.
  Benini}.} \bibinfo{year}{2019}\natexlab{}.
\newblock \showarticletitle{The Cost of Application-Class Processing: Energy
  and Performance Analysis of a {Linux}-Ready {1.7-GHz 64-Bit RISC-V} Core in
  {22-nm FDSOI} Technology}.
\newblock \bibinfo{journal}{\emph{IEEE Transactions on Very Large Scale
  Integration Systems}} \bibinfo{volume}{27}, \bibinfo{number}{11}
  (\bibinfo{date}{Nov.} \bibinfo{year}{2019}), \bibinfo{pages}{2629--2640}.
\newblock


\end{thebibliography}
}


\end{document}